\documentclass[sigconf]{acmart}

\usepackage[ruled,linesnumbered]{algorithm2e}    
\usepackage{amsmath}                             
\usepackage{amsthm}                              
\usepackage{balance}
\usepackage{booktabs}                            
\usepackage{centernot}                           
\usepackage{colortbl}                            
\usepackage{graphicx}                            
\usepackage{hhline}                              
\usepackage{lipsum}                              
\usepackage{mathtools}                           
\usepackage[all]{nowidow}                        
\usepackage{optidef}                             
\usepackage{pgfplotstable}                       
\usepackage{subcaption}                          
\usepackage{siunitx}                             
\usepackage{substr}                              
\usepackage{tikz}                                
\usepackage{tcolorbox}                           
\usepackage{xcolor}                              
\usepackage{hyperref}                            
\usepackage{xcolor}                              

\hypersetup{colorlinks, linkcolor=blue}
\usetikzlibrary{calc, fit, shapes.geometric}
\pgfplotsset{layers/layerss/.define layer set={
    bg,main,fg}{},set layers=layerss,}
\SetArgSty{textup}                               
\pgfplotsset{compat=1.16,
    compat/show suggested version=false}         
\DeclareSIUnit[number-unit-product=]\percent{
    \char`\%}

\definecolor{owhite}{rgb}{0.90, 0.90, 0.90}      
\definecolor{oblack}{rgb}{0.15, 0.15, 0.15}      

\newcommand{\ie}{i.e.,}                          
\newcommand{\eg}{e.g.,}                          
\newcommand{\etc}{etc.}                          

\DeclarePairedDelimiter\abs{\lvert}{\rvert}      
\DeclarePairedDelimiter\norm{\lVert}{\rVert}     
\DeclareMathOperator*{\argmax}{arg\,max\,}       
\DeclareMathOperator*{\sign}{sgn}                
\newcommand{\bigo}{\mathcal{O}}                  
\newcommand{\set}[1]{\{#1\}}                     
\newcommand{\powerset}{\mathscr{P}}              
\newcommand{\jacobian}{\mathbf{J}}               
\newcommand{\lp}[1]{\(\ell_{#1}\)}               
\newcommand{\ball}[2]{\mathcal{B}_#1(#2)}        

\newcommand{\shortsection}[2][.]{
    \vspace{1mm}\noindent\textbf{#2#1}}
\makeatletter\newenvironment{algo}[1][ht]{
    \def\@algocf@post@ruled{%
        \kern\interspacealgoruled\hrule%
        height\algoheightrule\kern3pt\relax}%
    \def\@algocf@capt@ruled{under}%
    \setlength\algotitleheightrule{0pt}%
    \SetAlgoCaptionLayout{centerline}%
    \begin{algorithm}[#1]}{%
    \end{algorithm}}
\makeatother

\newcommand{\jsma}{\texttt{JSMA}}                
\newcommand{\pgd}{\texttt{PGD}}                  
\newcommand{\cspfull}{\texttt{%
    Constrained Saliency Projection}}            
\newcommand{\csp}{\texttt{CSP}}                  
\newcommand{\dpllfull}{\texttt{%
    Davis-\-Putnam-\-Logemann-\-Loveland}}             
\newcommand{\dpll}{\texttt{DPLL}}                
\newcommand{\optics}{\texttt{OPTICS}}            
\newcommand{\ch}{\texttt{Cleverhans}}            
\newcommand{\pt}{\texttt{PyTorch}}               
\newcommand{\nsl}{\texttt{NSL-KDD}}              
\newcommand{\ph}{\texttt{Phishing}}              
\newcommand{\ddos}{\texttt{CICDDoS2019}}         
\newcommand{\drebin}{\texttt{DREBIN}}            
\newcommand{\aninea}{\texttt{a9a}}               

\AtBeginDocument{%
  \providecommand\BibTeX{{%
    \normalfont B\kern-0.5em{\scshape i\kern-0.25em b}\kern-0.8em\TeX}}}

\copyrightyear{2021}
\acmYear{2021}
\setcopyright{acmlicensed}
\acmConference[CCS '21]{Proceedings of the 2021 ACM SIGSAC Conference on Computer and Communications Security}{November 15--19, 2021}{Virtual Event, Republic of Korea}
\acmBooktitle{Proceedings of the 2021 ACM SIGSAC Conference on Computer and Communications Security (CCS '21), November 15--19, 2021, Virtual Event, Republic of Korea}
\acmPrice{15.00}
\acmDOI{10.1145/3460120.3484570}
\acmISBN{978-1-4503-8454-4/21/11}

\begin{document}
\fancyhead{}

\title{On the Robustness of Domain Constraints}

\author{Ryan Sheatsley}
\orcid{0000-0001-8447-602X}
\affiliation{%
    \institution{The Pennsylvania State University}
  \city{State College}
  \state{Pennsylvania}
  \country{USA}
}
\email{sheatsley@psu.edu}

\author{Blaine Hoak}
\affiliation{%
    \institution{The Pennsylvania State University}
  \city{State College}
  \state{Pennsylvania}
  \country{USA}
}
\email{bhoak@psu.edu}

\author{Eric Pauley}
\orcid{0000-0002-2197-9137}
\affiliation{%
    \institution{The Pennsylvania State University}
  \city{State College}
  \state{Pennsylvania}
  \country{USA}
}
\email{epauley@psu.edu}

\author{Yohan Beugin}
\affiliation{%
    \institution{The Pennsylvania State University}
  \city{State College}
  \state{Pennsylvania}
  \country{USA}
}
\email{ybeugin@psu.edu}

\author{Michael J. Weisman}
\affiliation{%
    \institution{United States Army Combat Capabilities Development Command Army Research Laboratory}
  \city{Adelphi}
  \state{Maryland}
  \country{USA}
}
\email{michael.j.weisman2.civ@army.mil}

\author{Patrick McDaniel}
\orcid{0000-0003-2091-7484}
\affiliation{%
    \institution{The Pennsylvania State University}
  \city{State College}
  \state{Pennsylvania}
  \country{USA}
}
\email{mcdaniel@cse.psu.edu}

\renewcommand{\shortauthors}{Sheatsley et al.}

\begin{abstract}\label{abstract}

    Machine learning is vulnerable to \textit{adversarial examples}--inputs
    designed to cause models to perform poorly. However, it is unclear if
    adversarial examples represent realistic inputs in the modeled domains.
    Diverse domains such as networks and phishing have \textit{domain
    constraints}--complex relationships between features that an adversary must
    satisfy for an attack to be realized (in addition to any adversary-specific
    goals). In this paper, we explore how domain constraints limit adversarial
    capabilities and how adversaries can adapt their strategies to create
    realistic (constraint-compliant) examples. In this, we develop techniques
    to learn domain constraints from data, and show how the learned constraints
    can be integrated into the adversarial crafting process. We evaluate the
    efficacy of our approach in network intrusion and phishing datasets and
    find: (1) up to \SI{82}{\percent} of adversarial examples produced by
    state-of-the-art crafting algorithms violate domain constraints, (2) domain
    constraints are robust to adversarial examples; enforcing
    constraints yields an increase in model accuracy by up to
    \SI{34}{\percent}. We observe not only that adversaries must alter inputs
    to satisfy domain constraints, but that these constraints make the
    generation of valid adversarial examples far more challenging.
    
\end{abstract}



\maketitle
\section{Introduction}\label{introduction}

Machine learning has demonstrated exceptional problem-solving capabilities; it
has become \textit{the} tool to learn, tune, and deploy for many important
domains, including healthcare, finance, education, and
security~\cite{esteva_guide_2019, heaton_deep_2016, popenici_exploring_2017,
buczak_survey_2016}. However, machine learning is not without its own
limitations: countless works have demonstrated the fragility of models when in
the presence of an adversary~\cite{goodfellow_explaining_2014,
kurakin_adversarial_2016, papernot_limitations_2016, carlini_towards_2017,
madry_towards_2017}. Across a variety of threat models, research has shown how
adversaries fully control model outputs, by classifying school buses as
ostriches~\cite{szegedy_intriguing_2013}, students as
celebrities~\cite{sharif_accessorize_2016}, or generating photos of synthetic,
yet unsettlingly realistic, people~\cite{goodfellow_generative_2014}.

This field of \textit{adversarial machine learning} is rich with research into
the exploitation of these models. While alarming to domains that have observed
significant advancements as a result of machine learning (\ie{} network
intrusion detection, spam, malware, \etc{}), it is not clear yet whether these
domains are as vulnerable as posited. This observation is rooted in the domain
which exemplifies the end-to-end capability of deep learning: images.
Influential works often use images as an empirical demonstration of their
findings~\cite{goodfellow_explaining_2014, papernot_limitations_2016,
carlini_towards_2017, madry_towards_2017, kurakin_adversarial_2016}. This has
an implicit assumption on the underlying threat models; adversaries can
manipulate features arbitrarily and independently. In other words, adversaries
are assumed to have \textit{full control} over the feature space and, more
importantly, \textit{all} input manipulations are equally permissible in the
domain under investigation. While often bound (exclusively) by some
self-imposed \lp{\mathscr{p}}-norm (canonically used as a surrogate for human
perception), there are constructs, rules, and other forms of \textit{domain
constraints} that many domains contain which images do not.

Domain constraints\footnotemark{} describe relationships between features. For
example, in network flow data, \textsc{TCP} flags can only be set for
\textsc{TCP} flows in networks---having these flags for \textsc{UDP} would
violate the semantics of the underlying phenomenon (network protocols).
Constraints encode the maneuvers (\ie{} perturbations) that are possible for an
adversary when crafting adversarial examples. Yet, existing threat models
broadly ignore this requirement, serving to generate examples that may or may
not represent legitimate examples of the domain. Thus, any vigilant system
could simply discard non-compliant samples because they are manifestly
adversarial---they do not represent a sample that could benignly exist. Such
detectable adversarial examples, thus, pose no risk.

\footnotetext{Note that the constraints discussed in this paper are different
from \textit{adversarial constraints}, which describe what an adversary seeks
to achieve (commonly, a classification mismatch between model and human).}

We argue in this paper that to properly assess the practical vulnerability of
machine learning in a domain, the constraints that characterize the domain must
be learned, incorporated, and demonstrated in attacks. Naturally, some domains
contain incredibly rigid structures (\eg{} binaries or networks) that could
offer robustness against adversaries (that is, an \textit{inability} to craft
realizable adversarial examples), which is one of the central questions we
investigate in this paper. By learning domain constraints, those who use
machine learning can build accurate threat models and thus, properly assess
realistic attack vectors.

Learning what the constraints are in arbitrary domains is a non-trivial
process; domains can contain multiple layers of complex abstractions,
frustrating any manual constraint identification through expertise.
Fortunately, there are data-driven approaches for learning and encoding useful
representations of constraints from areas within formal logic. One such method
comes from the seminal work of Leslie Valiant on PAC learning (probably
approximately correct learning)~\cite{valiant_theory_1984}. In this work,
Valiant described a setting for learning boolean constraints (specifically,
\(k\)-conjunctive normal form (CNF)) theories from data. Valiant's constraint
theory formulation and paired learning protocol provides a simple, yet
exhaustive, mechanism for identifying constraints and, coincidentally, an
elegant representation for integration into adversarial crafting algorithms.

In order to understand the robustness provided by domain constraints, we
characterize the \textit{worst-case adversary}. Specifically, the worst-case
adversary is defined as one who is \textit{least constrained}.  Said formally,
the number of possible observations rejected by a constraint theory is minimal.
We describe the worst-case adversary by exploiting a theoretical property in
our setting; a constraint theory is \textit{sound} if the observations it
certifies comply with the domain constraints. From this property, we show that
sound constraint theories reduce to memorization of the training data, and how
easing soundness yields generalization, with the worst-case adversary occurring
under the \textit{most general} (\ie{} least constrained) constraint theory
that can be learned.

In this paper, we explore adversarial examples with domain constraints by
answering two fundamental questions: (1) \textit{How would adversaries launch
attacks in constrained domains?}, and (2) \textit{Are constrained domains
robust against adversarial examples?} We design our approach by leveraging
frameworks within formal logic to learn constraints from data. Then, we modify
an algorithm for constraint satisfaction, \dpllfull{} (\dpll{}), to project
adversarial examples onto a constraint-compliant space. Finally, we introduce a
new adversarial crafting algorithm, the \cspfull{} (\csp{}): a blend of two
popular adversarial crafting algorithms that, by design, aids \dpll{} in
projection. We evaluate the efficacy of our approach on network intrusion
detection and phishing datasets. From our investigation, we argue that
incorporating domain constraints into threat models is \textit{necessary} to
produce realistic adversarial examples, and more importantly, constrained
domains are naturally more robust to adversarial examples than unconstrained
domains (\eg{} images).

\vspace{3pt}\noindent{}We contribute the following:
\begin{enumerate}

    \item We formalize \textit{domain constraints} in machine learning.

    \item We prove \textit{theoretical guarantees} for learning
        domain constraints in our setting. We describe when and why constraint
        theories are \textit{sound} and \textit{complete}, demonstrating an
        inherent trade-off between generalization and soundness.

    \item We satisfy domain constraints by \textit{projecting} non-realizable
        adversarial examples onto the space of valid inputs. We also introduce the
        \cspfull{} (\csp{}).

    \item We demonstrate the robustness produced by domain constraints against
        worst-case adversaries in two diverse datasets.  We observe that
        enforcing domain constraints can improve the robustness of a model
        substantially; in one experiment constraint enforcement restored model
        accuracy by \SI{34}{\percent}.

\end{enumerate}

\section{Overview}\label{overview}

\begin{figure}[t]
    \centering
    \includegraphics[width=\columnwidth]{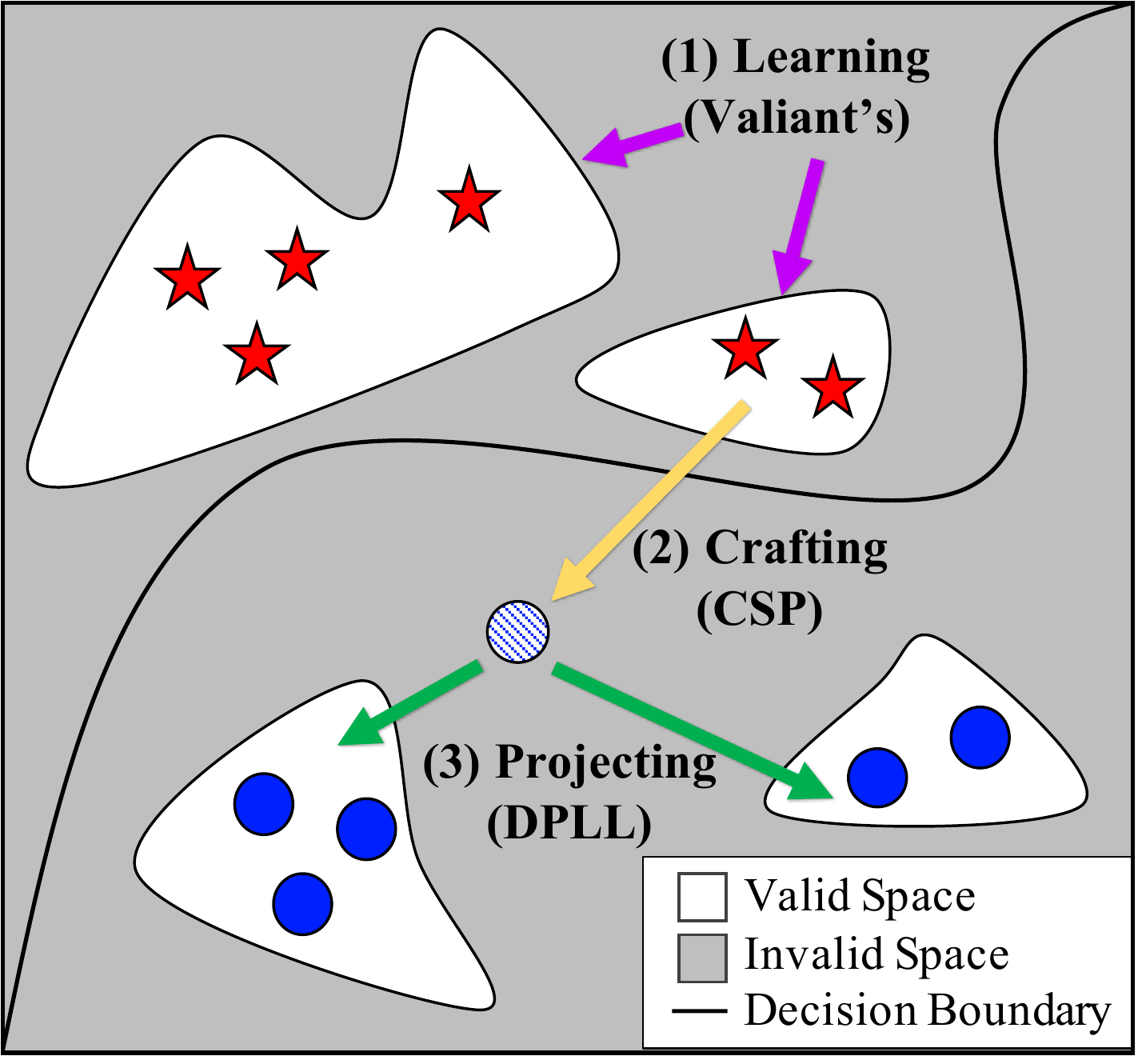}

    \vspace{-3mm} \caption{An Overview of Our Approach---Our approach consists
    of three steps: (1) learning the domain constraints, (2) crafting
    adversarial examples, and (3) projecting adversarial examples onto a
    constraint-compliant space.}\label{fig:overview}
\end{figure}

In order to measure the robustness of constrained domains against adversarial
examples, we must build a new set of techniques. We can envision this process
in three parts: (1) learning the domain constraints, (2) crafting adversarial
examples, and (3) projecting adversarial examples onto a constraint-compliant
space. A visualization of this approach is described in
Figure~\ref{fig:overview}.

\shortsection{Learning Constraints} In many domains, there are regions for
which samples do not exist (\eg{} \textsc{UDP} flows in the network domain do
not have \textsc{TCP} flags). The structures and rules that define these
regions may be complex and thus, we need a general approach to \textit{learn}
how domains are partitioned into valid and invalid regions. We leverage an
algorithm from PAC learning, Valiant's algorithm, to learn these domain
constraints in Section~\ref{approach:learning_constriants}.

\shortsection{Crafting Adversarial Examples} After domain constraints have been
learned, our approach can leverage \textit{any} adversarial crafting algorithm.
While there are many~\cite{goodfellow_explaining_2014,
papernot_limitations_2016,carlini_towards_2017,madry_towards_2017} attacks in
literature, we focus on \pgd{} in
Section~\ref{approach:crafting_adversarial_examples} as it is considered to be
the state-of-the-art for first-order adversaries. We will also consider a
constraint-specific approach in Section~\ref{appraoch:improving_projection}.

\shortsection{Projecting Adversarial Examples} With the learned domain
constraints and a set of adversarial examples, we must now enforce the
constraints on the crafted adversarial examples. We do this by
\textit{projecting} the adversarial examples onto the constraint-compliant
space (defined within some budget, as to preserve the goals of the adversary).
Specifically, we manipulate features in constraint-noncompliant adversarial
examples until they either satisfy the domain constraints or exceed the
allotted budget. We augment a seminal solver for constraint satisfaction,
\textsc{Davis-Putnam-Logeman-Loveland}, to perform this projection in
Section~\ref{approach:projecting_adversarial_examples}.

\section{Background}\label{background}

\subsection{Threat Model}\label{background:threat_model}

\shortsection{Adversarial Goals} Adversaries can have a variety of objectives,
from reducing model confidence to misclassifying a sample as a particular
target. Here, we focus on the former, that is, given a victim model
\(f_\theta\) with parameters \(\theta\), a sample \(e\), label \(y\), a
self-imposed budget \(\phi\) measured under some \lp{\mathscr{p}}-norm, and a
domain-dictated constraint theory \(T\) (which is our contribution), an
adversary aims to solve the following optimization objective:

\noindent\begin{argmini}|l|
    {\alpha}{\norm{\alpha}_{\mathscr{p}}}{}{}
    \addConstraint{f_\theta(e + \alpha) \neq{} y}
    \addConstraint{e + \alpha\in\ball{\phi}{e} \cap{} T}%
    \label{eq:ao}
\end{argmini}

\noindent Conceptually, the adversary searches within some norm-ball
\(\mathcal{B}\) of radius \(\phi\) around a sample \(e\) for a ``small'' change
\(\alpha\) to apply to \(e\) that yields the desirable model behavior (\ie{} an
\textit{adversarial example}). In the context of computer security, this could
translate to bypassing a network intrusion detection system.

``White-box'' settings represent the strongest adversaries and characterize
worst-case scenarios. Akin to insider threats, these adversaries have
unfettered queries to the model, can observe its parameters, and can use its
training data. From Equation~\ref{eq:ao}, suppose \(f_\theta\) describes the
model of the defender, then white-box adversaries either have direct access to
model parameters of \(f_\theta\) or the training data used to learn
\(f_\theta\).  Practically speaking, such adversaries can produce adversarial
examples with the tightest \lp{\mathscr{p}}-norm constraints (\ie{} the
smallest budgets). ``Gray-'' and ``Black-box'' threat models are two other
popular threat models that remove the degree to which an adversary has access
to model parameters or training data. These limited adversaries usually require
unique techniques for attacks to be
successful~\cite{papernot_transferability_2016,liu_delving_2016,
tramer_space_2017}.

In our work, we explore the efficacy of white-box adversaries when domain
constraints are enforced. Specifically, the adversary seeks to minimize model
\textit{accuracy} (that is, the number of samples correctly classified over the
total number of samples). Referring to Equation~\ref{eq:ao}, the adversary
computes samples classified as any class \(\hat{y}\neq y\) (where \(\hat{y}\)
is the prediction). Under this objective, the adversary attempts to reduce the
confidence an operator has in the predictions of a model.

\subsection{Adversarial Machine Learning}\label{background:AML}

We describe the two algorithms that serve as the basis for the algorithm used
in our approach discussed later in
Section~\ref{approach:crafting_adversarial_examples}.

\shortsection{Jacobian-based Saliency Map Approach (JSMA)} The
\jsma{}~\cite{papernot_limitations_2016} is an iterative approach leveraging
\textit{saliency maps}: a heuristic applied to the Jacobian matrix of the
model. For our work, the insightful component of the \jsma{} is that it selects
a single feature to perturb per iteration (\ie{} it optimizes over
\lp{0}-norms). This is especially important for security-critical domains as
other \lp{\mathscr{p} \neq{} 0}-norms are largely driven as surrogates for
human perception, and do not apply to the studied domains.

\shortsection{Projected Gradient Descent (PGD)} \pgd{}\footnotemark{}, by Madry
et al., is considered to be the state-of-the-art for first-order
adversaries~\cite{madry_towards_2017}.  Similar to the \texttt{Fast Gradient
Sign Method}~\cite{goodfellow_explaining_2014}, \pgd{} iteratively multiplies a
step-size \(\alpha\) by the sign of the gradient of the loss to perturb the
sample. As \pgd{} has been proposed as a ``universal adversary (among
first-order approaches),''~\cite{madry_towards_2017} we use it to evaluate the
robustness of constraints against adversarial examples.

\footnotetext{The ``projection'' in \pgd{} is unlike the ``projection'' used in
this work through \dpll{}.}

\section{Approach}\label{approach}

\subsection{Preliminaries}\label{approach:preliminaries} We overview constraint
learning and define the three algorithms we leverage for constraint learning,
projection, and clustering.

\shortsection{Problem Statement}~\cite{raedt_learning_nodate} provided one
formulation for framing constraint learning as a concept-learning problem. We
restate the relevant parts of the formalization for our work. Namely, we are
given a domain \(X=(X_1,\dots,X_n)\), where \(X_i\subseteq\mathbb{Z}\) (\ie{}
we have \(n\) features where \(X_n\) denotes the unique values feature \(n\)
can take), a space \(C\) of possible constraints (represented as \(k\)-CNF
Boolean formulae defined over \(X\)), and a set \(E\) of collected
observations. Our objective is to find a constraint theory \(T\) (\(T\subseteq
C\)) such that \(T\) certifies all observations \(e\in E\). We say \(T\)
\textit{certifies} \(e\) when all clauses \(t\in T\) are satisfied by \(e\). We
say a clause \(t\) is \textit{satisfied} when at least one literal in \(t\) is
\textsc{True} (where the features in \(e\) assign values to the literals in
\(t\)).  Conceptually, the collected observations \(E\) encode the structures
or rules of the domain (\ie{} \(T\)), that we seek to learn.

\shortsection{Valiant's Algorithm} Valiant's algorithm
(Algorithm~\ref{alg:valiant}) is a general, exhaustive, generate-and-test
algorithm for constraint learning~\cite{valiant_theory_1984}. The algorithm is
initialized with a constraint theory \(T\) of all possible constraints over
\(X\) (\ie{} \(T=C\)). Then, for each observation \(e \in E\), clauses that are
not satisfied by \(e\) are removed from \(T\). The algorithm terminates when
\(E\) has been exhaustively processed. Valiant's algorithm will converge to the
correct solution, assuming \(E\) sufficiently represents the domain and is free
of noise. Moreover, Valiant's algorithm does not require any \textit{negative
examples} (that is, known observations that violate domain constraints, which
are absent from popular machine learning datasets), which some constraint
learning approaches require~\cite{garg_ice_2014,zhu_data-driven_2018}.
Conceptually, \(T\) initially describes the \textit{space of possible
constraints} and, after applying \(T\) and the set, \(E\), of collected
observations to Valiant's algorithm, \(T\) will contain the set of constraints
that are satisfied by \textit{all} observations in \(E\) (in other words, the
intersection of satisfied constraints across the observations in \(E\)).

\begin{algo}[t]
    \SetAlgoLined{}
    \KwIn{observations \(E\), boolean clauses \(T\)}
    \KwOut{a conjunction of the remaining clauses \(T\)}
    \For{\(e\in E\)}{%
        \For{\(c\in T\)}{%
            remove \(c\) from \(T\) if \(e\nvdash c\)
        }
    }
    \Return{\(T\)}
    \caption{Valiant's Algorithm}\label{alg:valiant}
\end{algo}

\begin{algo}[t]
    \SetAlgoLined{}
    \SetKwFunction{KwRelax}{RelaxToCNF}
    \KwIn{a domain \(X\)}
    \KwOut{the space of possible constraints \(C\)}

    \(P\leftarrow\set{\powerset{}(X_i)\setminus\set{\varnothing, X_i}\mid X_i\in
    X}\)\\
    \(C\leftarrow\set{P_1\times\dotsb\times P_n}\), \(P_i\in P\)\\
    \(C \leftarrow\) \KwRelax{C} \\
    \Return{\(C\)}\\

    \caption{Generating the Space of Possible Constraints}\label{alg:gc}
\end{algo}

To illustrate how Valiant's algorithm operates, consider an example where a
dataset contains samples with two binary features, \(X_1=\set{x_1, \neg x_1}\)
and \(X_2=\set{x_2,\neg x_2}\). Then, \(C\), the space of possible constraints
is:

\noindent\begin{equation*}
    C = (x_1 \lor x_2)\land (x_1 \lor\neg x_2)\land (\neg x_1\lor x_2)\land
    (\neg x_1\lor\neg x_2)
\end{equation*}

\noindent If we initialize \(T\) (our target constraint theory) to \(C\), and
suppose \(E\) (our set of collected observations) consists of two observations
\(e_1 = (\textsc{False}, \textsc{False})\) and \(e_2 = (\textsc{True},
\textsc{True})\), then Valiant's algorithm first removes \((x_1 \lor x_2)\),
and then removes \((\neg x_1\lor\neg x_2)\) (as they are not satisfied by
\(e_1\) and \(e_2\), respectively). Our final constraint theory \(T\) is then
\((x_1\lor\neg x_2)\land (\neg x_1\lor x2)\). This example shows the thesis
behind Valiant's algorithm: \textit{only constraints that have support from
\textbf{all} observations are permissible.}

\shortsection{DPLL} For adversarial examples that do not comply with domain
constraints, we use an algorithm from the constraint satisfaction community,
\textsc{Davis-Putnam-Logeman-Loveland} (\dpll{})~\cite{davis_computing_1960},
to project adversarial examples onto the learned constraint theory returned by
Valiant's Algorithm. \dpll{} has some characteristics that make it ideal for
our task, namely: (1) it accepts boolean formulae in CNF (which is the native
form of the constraint theories learned by Valiant's Algorithm), and (2) it is
a \textit{backtracking-based} search algorithm: \dpll{} iteratively builds
candidate solutions for a given expression, which is a property we exploit,
detailed later in Section~\ref{approach:projecting_adversarial_examples}.
Further details about \dpll{} can be found in Appendix~\ref{appendix-c:dpll}

\shortsection{OPTICS} Later, we show how we model arbitrary data types as
domain constraints. To support this generalization, we leverage a clustering
algorithm, \textsc{Ordering Points to Identify the Clustering Structure}
(\optics{})~\cite{ankerst_optics_1999}. \optics{} has two advantages over other
clustering algorithms for our application, namely: (1) it scales to large
sample sizes, and (2) it is \textit{not} parameterized on specifying the number
of clusters. The second property is particularly important as parameterizing
the number of clusters assumes a priori knowledge of the constraints before
learning them in the first place.

\subsection{Learning Constraints}\label{approach:learning_constriants}

Recall our problem statement: given a domain \(X\), we first generate the space
\(C\) of possible constraints, then, with a dataset \(E\) of samples, we use
Valiant's algorithm to prune constraints that do not comply with samples in our
dataset (\ie{} \(T\subseteq C\)). Valiant's algorithm is elegant for binary
features and the fact that it produces ``hard'' boolean constraints makes it
attractive for encoding rigorous structures of domains. However, novel
applications of machine learning seldom use binary features exclusively;
categorical and continuous features (\eg{} packet rates or word counts) are
used in nearly every modern application of machine learning.

To address this limitation, there are two modifications that must be made: (1)
how the space of possible constraints is generated, and (2) how to determine if
a particular sample is certified by a constraint theory. We discuss these two
modifications below.

\shortsection{The Space of Possible Constraints} For boolean-only constraint
theories, the space of possible constraints is on the order of
\(\bigo{}(2^n)\), where \(n\) is the number of features. To account for
categorical features, we can further generalize this bound to
\(\bigo{}(2^{n\cdot i})\) by considering a one-hot encoding, where \(i\)
represents the largest cardinality of possible values among \(n\) features.
This guides us on not only how to generate the space of constraints, but also
how to modify Valiant's algorithm to accept a richer representation of
constraints.

Our approach, shown in Algorithm~\ref{alg:gc}, is as follows: first, given a
domain \(X\), we compute a pseudo-power set \(P_i\) from the set \(X_i\) of
unique values for feature \(i\). \(P_i\) is a pseudo-power set as we remove the
empty set (\ie{} no value is valid for the feature) and \(X_i\) (which would
allow the constraint to be trivially satisfied by any sample). We then perform
the Cartesian product over \(P\), which returns the set of all possible
combinations of constraints; this is the input to Valiant's algorithm (\(C\)).
At this stage, \(C\) contains sets \(P_i\) of sets \(p_i\), and so, we
transform this representation to CNF (\texttt{RelaxToCNF}) by adding
disjunctions between each \(p_i\in P_i\) and, finally adding conjunctions
between each \(P_i\in C\).

To adapt Valiant's algorithm to perform on a set-based representation of
constraints (\ie{} boolean \textit{and} categorical variables), we redefine the
\(\vdash\) operator (\ie{} logical entailment); instead of evaluating whether
or not a feature value causes a boolean assignment to be satisfied, we instead
evaluate if it is a member of the set (\ie{} \(\nvdash\) now operates as
\(\notin\), set membership, in Algorithm~\ref{alg:valiant}).

Consider the following example: suppose two features, \(X_1\) and \(X_2\). Let
\(X_1\) be a boolean variable (encoded as \(x_1\in\set{0, 1}\)) and let \(X_2\)
be a categorical variable that can take on one of three values (encoded as
\(x_2\in\set{A,B,C}\)). With our approach above, the space of possible
constraints is then:

\vspace{1pt}\noindent\begin{equation*}
    \begin{split}
        C = (x_1 \in \set{0} \lor x_2 \in \set{A}) & \land (x_1 \in \set{0} \lor x_2 \in \set{B}) \\
         & \cdots \\
         \land (x_1 \in \set{1} \lor x_2 \in \set{A, C}) & \land (x_1 \in \set{1} \lor x_2 \in \set{B, C})
    \end{split}
\end{equation*}\vspace{1pt}

\noindent If we initialize our target constraint theory \(T\) to \(C\), and let
\(E\) (our training data) consists of four samples \(e_1 = (0, A)\), \(e_2 =
(0, B)\), \(e_3 = (1, B)\), and \(e_4 = (1, C)\), then our objective is to
guide Valiant's algorithm to learn the following:

\noindent\begin{equation*}
    (x_1 \in\set{0}\land x_2 \in\set{A, B})\lor (x_1\in\set{1}\land x_2
    \in\set{B, C})
\end{equation*}

Conceptually, we can imagine a case where \(X_1\) describes a protocol (\eg{}
\texttt{TCP} or \texttt{UDP}), while \(X_2\) describes a service (\eg{}
\texttt{SSH}, \texttt{DNS}, or \texttt{NTP}). Then, our target constraint
theory \(T\) would be to learn that \texttt{SSH} can only be used with
\texttt{TCP}, \texttt{NTP} can only be used with \texttt{UDP}, while
\texttt{DNS} can be used with either \texttt{TCP} or \texttt{UDP}.

After running Valiant's algorithm on this example, our final constraint theory
\(T\) will be:

\noindent\begin{equation*}
    (x_1 \in\set{0}\lor x_2 \in\set{B, C})\land (x_1\in\set{1}\lor x_2
    \in\set{A, B})
\end{equation*}

Using the distributive law, we see that this is precisely what we sought to
learn. As a sanity check, we can see that the observations \((0, C)\) and \((1,
A)\) violate \(T\), which was our desired result.

\shortsection{Discretizing \(\mathbb{R}\)} While the framework above can
express a richer expression of constraints than boolean theories, it cannot
model variables that live within the domain of real numbers \(\mathbb{R}\).
However, modeling constraints that have inequalities such as \(\set{x \mid 0.25
\leq x \leq 0.80}\) is non-trivial, as inferring the proper ranges for a given
variable has no straightforward answer.

We are motivated to extend our set-based formulation of constraints to model
continuous variables, as our generalization from the boolean domain
\(\mathbb{B}\) to the domain of integers \(\mathbb{Z}\) has some ideal
properties: (1) the set-based formulation can model constraints that are
readily interpretable, (2) constraint-certification reduces to simple
membership tests, (3) elegant integration into constraint learning algorithms,
and, (4) gives a simple asymptotic bound to conceptualize the space of possible
constraints. These properties are attractive and later we will show how our
formulation can be integrated into adversarial crafting algorithms.

We leverage \optics{} to enable encoding of continuous features as discrete
values. Here we express constraints with \textit{sets of ranges} (\eg{}
\(\set{x\mid(0.25 \leq x_i < 0.50) \lor (0.75 \leq x_i < 1.00)}\)).
Specifically, continuous features in samples are mapped to the bins and
thereafter the associated constraints are learned as any other discrete
feature. Later, when we project adversarial examples (discussed in
Section~\ref{approach:projecting_adversarial_examples}), continuous features
have their values set to the edge value closest to the origin of the perturbed
value (\ie{} a value that is projected from a higher number is set to the top
of the bin range, and a lower number is set to the bottom of the range).

\subsection{Theoretical Guarantees}\label{approach:theorectical_guarantees}

In this section we define the properties that the learning process described
above guarantees. When the full space of possible constraints is considered,
our approach learns a constraint theory that is \textit{sound} with respect to
observations and domain constraints.  Soundness guarantees that if a sample is
certified by the constraint theory, then the sample complies with the domain
constraints. The approach yields a sound constraint theory through
Algorithm~\ref{alg:gc}, generating the space of possible constraints
(specifically the generation of the pseudo-power set). We first briefly
describe our principal findings and later discuss formally when a constraint
theory learned with our approach is sound and why.

Without loss of generality, consider that for all \(X_i\in X\), \(|X_i|=n\),
that is, the number of unique values for all features is \(n\). Then, the
pseudo-power set\footnotemark{} contains sets of cardinality
\(1,\dots,k,\dots,(n-1)\). The clauses learned from sets of cardinality 1
(\ie{} the cardinality of the literals in the clause is 1) represent the
\textit{most general} constraints, while clauses learned from sets of
cardinality \(n-1\) represent the \textit{least general} constraints (we show
later that they represent rote memorization of the training data).

\footnotetext{Recall that we generate a pseudo-power set by excluding the empty
set \(\emptyset\) and the feature space itself, which correspond to sets of
cardinality 0 and \(n\), respectively.}

Let \(k\) bound the maximum cardinality considered when generating the
pseudo-power set. Then, from our observations we gather that: (1) if \(k=n-1\)
(\ie{} the clauses generated contain literals of cardinality at most \(n-1\)),
the learned constraint theory is guaranteed to be sound, (2) if \(k=1\), the
learned constraint theory is \textit{maximally general}, and (3) for some \(k\)
in-between 1 and \(n\), there is a trade-off between the degree to which the
learned constraint theory is sound (with respect to domain constraints at
cardinality \(k\)) and how well it generalizes to unseen observations. Thus,
\(k\) allows us to ease soundness and gain generalization, which we exploit in
characterizing the worst-case adversary. In this way, \(k\) is the parameter
that is used to tune the learned constraint theory from general to sound.

Next, we formally define the theoretical properties of our approach and show
when and why they hold. Consider the following properties with respect to a
constraint theory \(T\) and observations \(e\):

\begin{enumerate}

    \item \textbf{sound}: if \(T\) certifies \(e\), then \(e\) complies with the
        domain constraints.

    \item \textbf{complete}: for all possible observations \(e\) that comply with
        the domain constraints, \(T\) certifies \(e\).

\end{enumerate}

\noindent Recall (Section~\ref{approach:preliminaries}) that \(T\)
\textit{certifies} \(e\) when all clauses \(t\in T\) are satisfied by \(e\). A
clause \(t\) is \textit{satisfied} when at least one literal in \(t\) is
\textsc{True} (where the features in \(e\) assign values to the literals in
\(t\)). Now, let \(\Xi\) represent the space of all possible observations. We
can partition \(\Xi\) into two subspaces, \(\Lambda\) (the space of
observations that comply with the domain constraints)  and \(\Psi\) (the space
of observations that do not comply with the domain constraints).  Clearly,
\(\Lambda\cup\Psi=\Xi\) and \(\Lambda\cap\Psi=\emptyset\). \(T\) is sound if it
does not certify any observations from \(\Psi\) and \(T\) is complete if it
certifies all observations from \(\Lambda\).

\shortsection[?]{When is \(T\) Complete} From the definition of complete, we
can gather that \(T\) is axiomatically complete when \(T\) arbitrarily
certifies \textit{any} observation. Said alternatively, given that \(T\) is
\textit{not} complete if it does not certify all observations from \(\Lambda\),
\(T\) can be axiomatically complete when it certifies all observations,
regardless if they come from \(\Lambda\) or \(\Psi\). For example, the empty
constraint theory \(T=\emptyset\) is complete, as it will certify all possible
observations \(e\in\Lambda\) that comply with the domain constraints (as well
as those that do not, \ie{} \(e\in\Psi\)).

\shortsection[?]{When is \(T\) Sound} From the definition of sound, we can
gather that \(T\) is axiomatically sound when \(T\) does not certify
\textit{any} observations. Said differently, given that \(T\) is \textit{not}
sound if \(T\) certifies a single observation from \(\Psi\), \(T\) can be
axiomatically sound when it refuses to certify any observation. For example,
when \(T\) equals the space of possible constraints \(C\), \(T\) is
axiomatically sound, as it will reject all possible observations \(e\in\Psi\)
that do not comply with the true domain constraints (as well as those that do
comply, \ie{} \(e\in\Lambda\)).

\shortsection{Properties of Our Approach} With the two settings for when \(T\)
is sound or complete, we now turn to the setting investigated in this paper.
Specifically, we highlight some facts: (1) we have a set of observations \(E\)
for which we know comply with the domain constraints (\ie{}
\(E\subseteq\Lambda\)), (2) we have no observations that do not comply with the
domain constraints (\ie{} \(E\cap\Psi=\emptyset\)), (3) Valiant's algorithm can
be initialized with the space of possible constraints (\ie{} \(T=C\)), which
entails: \(T\) \textit{initially contains a superset of the domain
constraints}. From (1), (3), and the fact that Valiant's algorithm returns the
intersection of satisfied constraints across the observations in \(E\), we can
derive the following: when \(T\) is initialized to \(C\), \textit{the learned
constraint theory returned by Valiant's algorithm is a superset of the domain
constraints}.

Notably, the degree to which \(T\) remains a superset (and not equal to) of the
domain constraints is a function of the quality of \(E\) in characterizing the
underlying phenomena; as \(E\) approaches \(\Lambda\), \(T\) converges on the
domain constraints. In this way, Valiant's algorithm prioritizes being sound
over being complete.

Concretely, for any amount of observations in \(E\), if \(T\) certifies a new
observation, then it complies with the domain constraints (as \(T\) contains a
superset of the domain constraints). However, if \(T\) does not certify a new
observation that \textit{does}, in fact, comply with the domain constraints, it
is because this new observation failed to satisfy a clause \(t^*\in T\) that
should have been removed. Valiant's algorithm would fail to remove the
erroneous clause \(t^*\) if \(E\) did not contain a counter-example for \(t^*\)
when learning \(T\). Thus, unless \(E=\Lambda\), \(T\) will not be complete
(but \(T\) will always be sound).

\shortsection[?]{Why is \(T\) Sound} The final piece in characterizing the
worst-case adversary in our setting is rooted in analyzing what makes \(T\)
sound. Specifically, \(T\) is sound through Algorithm~\ref{alg:gc}: Generating
the Space of Possible Constraints.  Recall, we compute the Cartesian product of
the pseudo-power set of unique values observed across all features. For each
feature, the generated pseudo-power set can be decomposed as the union of the
unique combinations of sets with cardinality \(1,\dots,k,\dots,(n-1)\), where
\(n\) represents the number of unique values observed for some feature \(i\)
(\ie{} \(|X_i|=n\)). For simplicity, consider only the clauses whose sets have
cardinality \(n-1\). Trivially, this means that such clauses include all values
for a given attribute, except one. Call this set of clauses \(C^*\). In this
setting, when \(E\) and \(C^*\) are passed into Valiant's algorithm,
\textit{Valiant's algorithm will return a learned theory \(T^*\) that is an
exclusive encoding of \(E\) }. In other words, \(T^*\) will \textit{only}
certify \(E\) and nothing else (we formally prove this in
Appendix~\ref{appendix-a}).

Note that, this is a useful fact as if \(E=\Lambda\), then it would be
desirable to learn a constraint theory that certified \(E\) and only \(E\).
However, when \(E\subset\Lambda\) (as ostensibly all practical applications of
machine learning do), then this reduces \textit{learning} to
\textit{memorization} (this is analogous to \textit{overfitting} in machine
learning). From a learning perspective, this encourages us to bound the
cardinality of pseudo-power set to be no greater than \(k\), such that the
learned constraint theory \textit{generalizes}. Moreover, from an adversarial
perspective, a constraint theory that \textit{memorizes} the training data
would require the adversary to craft adversarial examples that are precisely
the training data itself. For any well-trained model, this means we already
know where ``adversarial examples'' can exist: where the model produces errors
on the training set. In other words, this characterizes the \textit{best-case
adversary}.

\shortsection{The Worst-Case Adversary} With these facts, we now characterize
the \textit{worst-case} adversary. Above, we observed how computing clauses
with cardinalities of \(n-1\) results in constraint theories that certify the
training data exclusively (that is, the total number of possible observations
certified is at most \(|E|\)), which characterizes the \textit{best-case
adversary}. Thus, the \textit{worst-case adversary} in our setting is one where
the learned constraint theory is \textit{most general}, in other words,
certifies the \textit{maximal} number of observations (which subsequently
translates to certifying the maximal number of adversarial examples). To learn
a theory that certifies the maximal number of observations, we set \(k=1\) when
generating the pseudo-power set, which then results in clauses whose literals
have cardinality 1 (we formally prove such constraint theories certify a
maximal number of observations in Appendix~\ref{appendix-a}).

\subsection{Crafting Adversarial Examples}\label{approach:crafting_adversarial_examples}

\shortsection{On \lp{\mathscr{p}}} \lp{\mathscr{p}}-norms have been adopted by
the academic community as the de facto standard for measuring a form of
``adversarial constraints.'' That is, it serves as a measurement of
detectability as a surrogate for human perception (for image applications) or
some arbitrary limitation on adversarial capabilities. For the former, it has
been generally agreed upon that \lp{2} or \lp{\infty} serve as better estimates
of human perception. For the latter, adversarial capabilities are usually
argued from a domain-specific perspective. For non-visual domains, we argue
that the \lp{0} norm is most representative of adversarial capabilities for two
reasons: (1) distance across features in non-image data is not uniform;
\lp{\mathscr{p} \neq{} 0}-norms on varying data types and semantics bear no
meaningful interpretation, and (2) for non-image domains, the degree to which
an adversary can manipulate \textit{every} feature yields little insight versus
\textit{what} features an adversary can manipulate.

\shortsection{Adversarial Constraints} Yet another important topic of
discussion for applications of adversarial machine learning outside of images
is: \textit{what is the adversary trying to accomplish?} For images, this has
been rooted in the use of \lp{\mathscr{p}}-norms: there should be a
misclassification between human and machine. For other domains, each have their
own answer, \eg{} consumer reviews should be read as containing positive
sentiment by humans, yet classified as negative sentiment by machine (or
vice-versa)~\cite{papernot_crafting_2016}; malware should maintain its
malicious behavior, post-perturbation~\cite{grosse_adversarial_2017,
kolosnjaji_adversarial_2018}; speech recognition systems should incorrectly map
audio to commands versus what a human would hear~\cite{carlini_audio_2018},
among other objectives. For our work, we follow similar intuitive objectives,
that is, post-perturbation: malicious network flows must maintain their attack
goals and phishing websites must mimic victim websites.

After defining \textit{what} the goals of an adversary are, the next question
is: \textit{how do we know the adversary has met those goals?} For images and
text, it has assumed to be self-evident; peers can inspect images produced by a
crafting algorithm or read the altered text of a consumer review.  While
human-based verification is possible in some domains, in others (particularly
those that are security-critical) it is not. For example, to validate that a
network flow or a malware executable is malicious, then it must be replayed and
its behavior observed in its respective domain. However, this may not always be
possible; mapping back from a feature vector of an adversarial example to its
original form may be non-trivial or outright impossible (network intrusion
detection datasets may not always provide the packet captures used to build the
dataset). Therefore, we need an approach that preserves the goals of an
adversary when the original form of a sample cannot be rebuilt.

\begin{figure}[t]
    \centering
    \resizebox{\columnwidth}{!}{\input{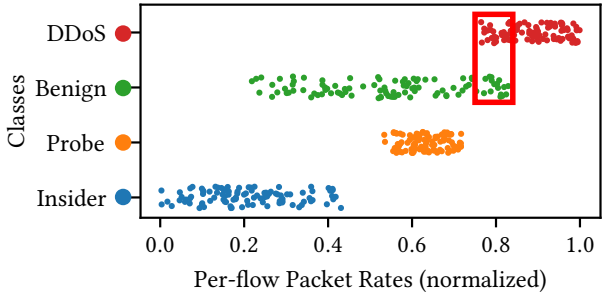}}
    \vspace{-6mm}\caption{Encoding Adversarial Constraints --- Distribution for
    normalized, per-flow \textsc{Packet Rates} with classes from the \nsl{}
    dataset. The region for the adversary attempting to classify \textsc{DDoS}
    traffic as \textsc{Benign} traffic, \textit{while maintaining the semantics
    of the attack}, is shown in the rectangle.}\label{fig:ac}
\end{figure}

To address this limitation, we formulate and add an additional layer of
``adversarial constraints'' that encode class-specific behaviors. Specifically,
we capture class-specific \textit{feature bounds}, which are either: (1) the
set of observations for categorical variables, or (2) the minima and maxima for
continuous variables. We argue that the underlying behavior of a sample is
preserved if it does not step outside of its bounds (which we hand validate).
The intuition is straightforward: by staying within these bounds, then we
produce adversarial examples with behaviors (as defined by feature values) that
\textit{have already been observed} for a particular class.

To illustrate this approach, consider Figure~\ref{fig:ac}. Shown in this
network intrusion detection example, there are four classes, \texttt{DDoS},
\texttt{Benign}, \texttt{Probe}, and \texttt{Insider} and each class has unique
range of values for the \texttt{Packet Rates} feature.  Suppose an adversary
begins with a malicious \texttt{DDoS} sample and wishes to have it
misclassified as \texttt{Benign}. Then, to preserve the underlying behavior of
the sample, any perturbation to the \texttt{DDoS} sample must be between
\(\sim\) \SI{0.75} and \(\sim\) \SI{1.0}. Naturally, the target region for the
adversary is shown in the box, where the packet rates for \texttt{DDoS} and
\texttt{Benign} overlap. We enforce adversarial constraints while generating
adversarial examples as well as projecting them.

\subsection{Projecting Adversarial Examples}\label{approach:projecting_adversarial_examples}

The final step in our approach is to project some adversarial example \(e^*\)
onto the constraint-compliant space described by the learned constraint theory
\(T\), as, if \(T\) does not certify \(e^*\), then \(e^*\) is not adversarial
at all, as it would not be \textit{realizable} (as \(T\) encodes what is valid
for a domain).  This projection is non-trivial as there could be multiple
features that do not comply with \(T\) and so, deciding which features to
modify in \(e^*\) so that \(T\) certifies \(e^*\) could induce other features
to become non-compliant, \etc{} Therefore, we need a mechanism that can
efficiently project \(e^*\) as to comply with \(T\).

This problem is isomorphic to constraint satisfaction problems; given a boolean
expression \(T\) with some number of clauses \(t\in T\), we seek to find an
assignment \(\set{\textsc{True}, \textsc{False}}\) to each literal in \(t\)
such that \(T\) is \textsc{True}. For our work, we use the \dpllfull{}
(\dpll{}) algorithm~\cite{davis_computing_1960}. As discussed earlier in
Section~\ref{approach:preliminaries}, \dpll{} has some properties that make it
ideal for our task, particularly that it is a \textit{backtracking-based}
search algorithm.

\dpll{} is parameterized on \(H\) (shown in Algorithm~\ref{alg:dpll} in
Appendix~\ref{appendix-c:dpll}), a boolean theory with \textit{partial} literal
assignments (adapted to accept our set-based constraint representation,
discussed in Section~\ref{approach:learning_constriants}).  We can exploit this
fact by projecting an adversarial example \(e^*\) onto the space described by
the learned constraint theory \(T\). Specifically, for each clause \(t\) in
\(T\), we determine which clauses are satisfied with respect to value of
features in \(e^*\). If all clauses \(t\in T\) are satisfied, then we simply
return the adversarial example (as \(T\) certifies \(e^*\)). Otherwise, for
each feature \(i\) in \(e^*\), we record the number of clauses \(t\in T\)
satisfied by \(e_i^*\).

Finally, we build \(H\) by allowing the bottom \(\phi\si{\percent}\) of
literals to be unassigned while we assign the top \(100-\phi\si{\percent}\) of
literals to the corresponding feature values in \(e^*\). Here, \(\phi\)
parameterizes: (1) the depth of the tree produced by \dpll{} (and thus its
runtime), and (2) indirectly controls the likelihood \(e^*\) will still be
misclassified (and thus, an adversarial example) after the assignment returned
by \dpll{} is applied to \(e^*\). To achieve (2), \(\phi\) should be small to
maintain the misclassification of \(e^*\), yet large enough that \dpll{} has a
sufficiently-sized search space to successfully project \(e^*\) onto \(T\).

\subsection{Improving Projection}\label{appraoch:improving_projection}

In our evaluation, we find that adversarial examples produced by \pgd{} often
fail to be projected onto the constraint-compliant space described by \(T\)
(within the allotted \(\phi\) budget). Our hypothesis on why these projections
failed stems from the fact that \pgd{} optimizes over the \lp{\infty} norm,
while the structure of the constraints and the budget used by \dpll{} may favor
adversarial crafting algorithms that optimize over the \lp{0} norm.  With this
hypothesis, we introduce our own \lp{0}-based adversarial crafting algorithm
that blends the iterative optimization of \pgd{} with \textit{saliency maps}
from the \jsma{}.

\shortsection{The Constrained-Saliency Projection (CSP)} The \csp{} is our
approach. Like \pgd{}, we consider a powerful adversary who can take multiple
steps on the sign of the gradient of some loss function (or, in our case, the
Jacobian of the model) and like the \jsma{}, the adversary computes
\textit{saliency maps} to determine the single most influential feature (and
thus the feature to perturb), and unlike either, we project back onto a
constraint-compliant space (described by \(T\), our extracted constraint
theory). Formally, we define the \csp{} as:

\noindent\begin{equation*}
    \begin{split}
        \textbf{S} &= \texttt{SaliencyMap}\left(\hat{y}, \jacobian\left(e^{r}\right)\right) \\
        i &= \argmax_j \abs{S_{j}} \\
        e^{r+1}_i & = e^{r}_i +
        \alpha\cdot\sign S_{i} \\
    \end{split}
    \label{eq:csp}
\end{equation*}

\noindent where \(\textbf{S}\) is the saliency map for a target\footnotemark{}
class \(\hat{y}\), \(\jacobian\) is the Jacobian of a model with respect to the
\(k\)-th perturbation of a sample \(e\), \(i\) is a feature index, and
\(\alpha\) is the perturbation magnitude. We slightly tweak the definition for
\texttt{SaliencyMap} that is originally proposed
in~\cite{papernot_limitations_2016}:

\footnotetext{We also consider an untargeted variant of the \csp{} where we set
\(\hat{y}\) to the label \(y\) for sample \(e\) and use the negative of the
Jacobian.}

\noindent\begin{equation*}
    \texttt{SaliencyMap}_{i}\left(\hat{y}, \jacobian\right) =
    \begin{cases}
        0 & \textrm{if} \sign J_{\hat{y},i}=\sign (\sum\limits_{j\neq \hat{y}}J_{j,i})
        \\ J_{\hat{y},i} \cdot
        \abs{\sum\limits_{j\neq \hat{y}}J_{j,i}} & \textrm{otherwise} \\
    \end{cases}
    \label{eq:sm}
\end{equation*}

\noindent where \(J_{j,i}\) refers to the \(j\)th class and \(i\)th feature in
the model Jacobian. This approach yields a subtle improvement; the formulation
of the \jsma{} in~\cite{papernot_limitations_2016} required a perturbation
parameter that could be either positive or negative (which determined the
heuristic used to build the saliency maps). However, our formulation for
saliency maps allows the \csp{} to iteratively add or subtract from a feature
\(i\), depending on whichever is more advantageous for the adversary.

\section{Evaluation}\label{evaluation}

With our techniques to learn and integrate domain constraints into the
adversarial crafting process, we evaluate our approach on two diverse datasets.
We ask the following:

\begin{enumerate}

    \item Do known crafting algorithms violate domain constraints?

    \item Do domain constraints provide robustness?

\end{enumerate}

\subsection{Experimental Setup \& Datasets}\label{evaluation:setup}

Our experiments were performed on a Dell Precision T7600 with an Intel Xeon
E5--2630 and a NVIDIA Geforce TITAN X. We used
\ch{}~\cite{papernot_cleverhans_2016} for adversarial attacks and
\pt{}~\cite{paszke_pytorch_2019} for building models. We defer to
Appendix~\ref{appendix-c} for hyperparameters, architectures, and other
miscellanea concerning our models. The experimental datasets are summarized in
Table~\ref{tab:ds} and described below.

In the following figures, \csp{} and \pgd{} refer to the attacks
pre-projection, while the \texttt{Constrained-\(\cdot\)} variants show the
results, post-projection, with \dpll{} and the learned domain constraints.  We
report the rate of invalid samples as the number of adversarial examples that
do not comply with domain constraints over the total number of adversarial
examples crafted. Model accuracy is measured as the number of adversarial
examples classified correctly by the model over the total number of adversarial
examples crafted. For constrained variants, samples that do not comply with
domain constraints after projection are counted as correctly classified.

\begin{table}
    \begin{tabular}{lccc}
        \toprule
        Dataset & \# Samples & \# Features & \# Classes \\
        \midrule
        \nsl{} & \(\approx 10^5\) & 11  & 5 \\
        \ph{}  & \(\approx 10^4\) & 10 & 2 \\
        \bottomrule
    \end{tabular}
    \caption{Summary of Dataset Statistics\vspace{-.5cm}}\label{tab:ds}
\end{table}

\shortsection{\nsl{}} The \nsl{}~\cite{tavallaee_detailed_2009} is a subset of
the seminal \texttt{KDD Cup '99} network intrusion detection dataset, dating
back to 1999.  While the NIDS data is somewhat dated, the breath and depth of
the \nsl{} makes it ideal for studying the effect of domain constraints. The
dataset contains 125,973 samples for training and 22,544 for testing,
representing four attacks and benign traffic.

As~\cite{al-jarrah_machine-learning-based_2014} demonstrates, many features in
the \nsl{} describe redundant information. Thus, we apply the same feature
reduction techniques to the data to bring the feature space from 41 features
down to 11.  This had a minor impact on the accuracy of our models (\ie{} from
\SI{77}{\percent} down to \SI{70}{\percent}), yet drastically improved the
scalability of learning domain constraints from the data.

\shortsection{\ph{}} \ph{}~\cite{chiew_new_2019} is a dataset for identifying
website phishing. The dataset contains \textit{internal} and \textit{external}
features of a website, \eg{} \texttt{HTML} versus \texttt{WHOIS} records. The
features were extracted from 5,000 popular phishing websites and 5,000
legitimate webpages.

Moreover,~\cite{chiew_new_2019} demonstrates that, among the original 48
features, only ten were necessary to maximize model accuracy. We use these ten
suggested features to apply our constraint learning algorithms on and build our
models from.  As~\cite{chiew_new_2019} claimed, we were able to achieve maximal
model accuracy (\ie{} above \SI{94}{\percent}) with ten features.

\subsection{Learning Domain Constraints}\label{evaluation:learning_domain_constraints}

\shortsection{Constraint Representation} Modern machine learning datasets often
contain tens of thousands of samples and learned constraint theories can be
equally as large (or even greater for some domains). Thus the runtime
performance of the constraint generation and evaluation algorithms is
important; we use a representation of domain constraints so that evaluation is
efficient. Details on these optimizations are in
Appendix~\ref{appendix-c:constraint_representation}.

\shortsection{The Space of Constraints} Motivated by our theoretical analysis
in Section~\ref{approach:theorectical_guarantees}, our evaluation characterizes
the \textit{worst-case adversary}. That is, our learned constraint theory \(T\)
is maximally general in that it certifies the maximal number of observations
(\ie{} the threat surface of potential adversarial examples is maximal). To
this end, we bound the pseudo-power set \(P\), and therefore the literals in
each clause, to have cardinality  \(k=1\).

\shortsection{Constraints Learned From Our Datasets} After having generated
clauses whose literals have cardinality \(k=1\), we pass \(C\) and the full
datasets into Valiant's algorithm. We find that the \nsl{}, the network
intrusion detection dataset, contained the most constraints at 5,330, while
only 1,995 constraints were learned from \ph{}, the phishing websites dataset.
We will provide some introspection on the constraints learned from the \nsl{}
later in Section~\ref{discussion}.

\subsection{Crafting Adversarial Examples}\label{evaluation:crafting_adversarial_examples}

For each dataset, we generate adversarial examples through both \pgd{} and the
\csp{}.  Both algorithms apply a 0.01 \lp{\infty} perturbation at each
iteration (\eg{} 35 iterations corresponds to a perturbation magnitude no
greater than 0.35, roughly a third of the feature space) to continuous
features. Perturbations to binary or categorical features are enforced to be -1
or 1 through one-hot encoding (as adversarial examples that report using 0.5
\texttt{TCP} for a \texttt{Protocol} feature are nonsensical). We compute
results by generating adversarial examples over the test set and measure
robustness through model accuracy.

\subsection{Projecting Adversarial Examples}\label{evaluation:projecting_adversarial_examples}

\shortsection{Selecting Features for \dpll{} to Perturb} Recall from
Section~\ref{approach:projecting_adversarial_examples}, we first identify the
feature values of adversarial examples that satisfy the fewest constraints
(therein identifying the features that are most constrained by the domain).
Intuitively, the heuristic we describe below is based on the following insight:
if the most constrained features are perturbed, then the resultant adversarial
example is likely to be rejected by the constraint theory. Thus, for each
adversarial example, we identify the most constrained features and use \dpll{}
to modify these features so that the sample is likely to be certified by the
learned constraint theory.

\begin{figure}[t]
    \centering
    \begin{subfigure}[t]{\columnwidth}
        \centering
        \resizebox{\columnwidth}{!}{\input{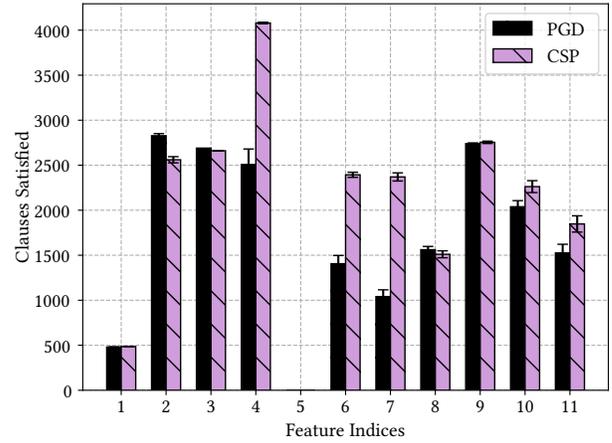}}
        \caption{\nsl{}}
    \end{subfigure}
    \begin{subfigure}[t]{\columnwidth}
        \centering
        \resizebox{\columnwidth}{!}{\input{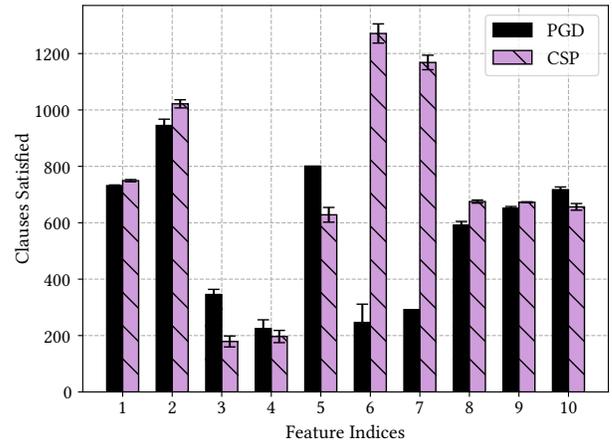}}
        \caption{\ph{}}
    \end{subfigure}
    \caption{Clause Satisfaction Bar Chart---The mean number of clauses
    satisfied from the learned constraint theories on a per-feature basis for
    the adversarial examples produced by \csp{} and \pgd{}. Error bars
    represent \SI{95}{\percent} C.I.}\label{fig:csh}
\end{figure}

Figure~\ref{fig:csh} shows clause satisfaction bar charts for \ph{} and the
\nsl{} after 35 iterations of perturbations by the \csp{} and \pgd{}.  Recall
that to use \dpll{} effectively, we wish to project constraint-noncompliant
samples onto the constraint-compliant space with minimal sample modification
under a \(\phi\) budget. Additional information on the features can be found in
Appendix~\ref{appendix-c:dataset_details}.

As the charts show, there are some features that satisfy significantly more
clauses than others across the majority of samples. Because \dpll{} is
parameterized on some allotted \(\phi\), this discrepancy suggests that we
might use clause satisfaction as a heuristic to select features for \dpll{} to
prioritize. In other words, we configure \dpll{} to prioritize projecting
features whose values satisfy few clauses (as opposed to features whose values
readily satisfy many clauses). We parameterize \dpll{} with an additional
\(\phi\) budget of \SI{20}{\percent} (\ie{} \dpll{} will select no greater than
the bottom \SI{20}{\percent} of features, as determined by the clause
satisfaction bar charts, to project adversarial examples onto the learned
constraint theory).

While this clause satisfaction heuristic is effective, other heuristics may
improve \dpll{} success further. For instance, high variance in number of
clauses satisfied by a feature across many samples could imply that the feature
is highly salient towards constraints. We defer investigation of additional
heuristics for future work.

\begin{figure*}[t]
    \centering
    \begin{subfigure}[t]{\columnwidth}
        \centering
        \resizebox{\columnwidth}{!}{\input{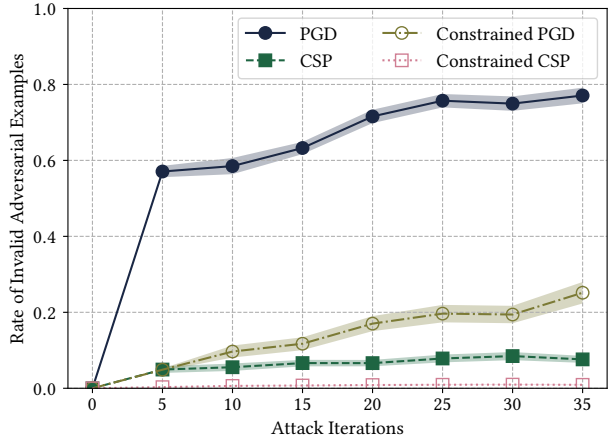}}
        \caption{\nsl{} --- Rate of Invalid Adversarial Examples}\label{fig:results:nsl:rate}
    \end{subfigure}
    \begin{subfigure}[t]{\columnwidth}
        \centering
         \resizebox{\columnwidth}{!}{\input{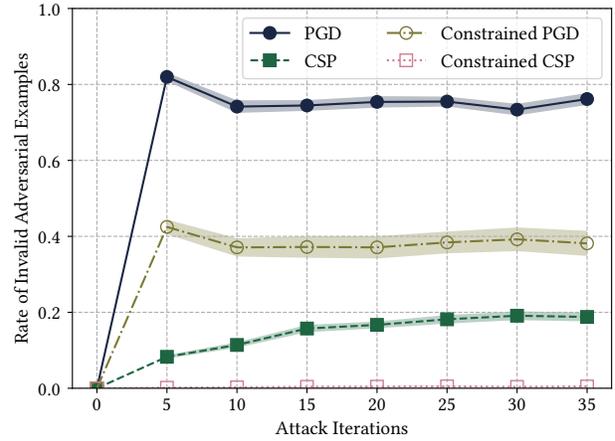}}
        \caption{\ph{} --- Rate of Invalid Adversarial Examples}\label{fig:results:phishing:rate}
    \end{subfigure}
    \medskip
    \begin{subfigure}[t]{\columnwidth}
        \centering
        \resizebox{\columnwidth}{!}{\input{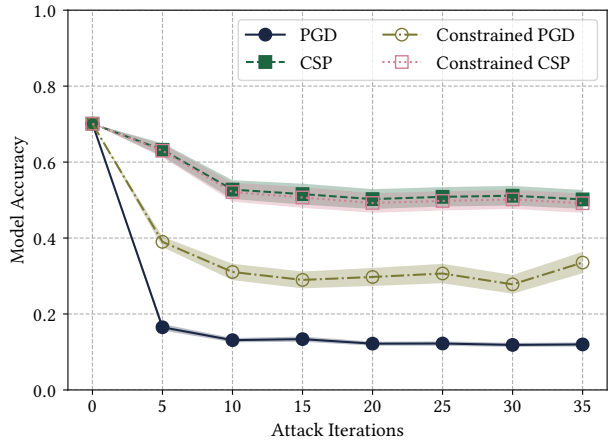}}
        \caption{\nsl{} --- Model Accuracy}\label{fig:results:nsl:accuracy}
    \end{subfigure}
    \begin{subfigure}[t]{\columnwidth}
        \centering
        \resizebox{\columnwidth}{!}{\input{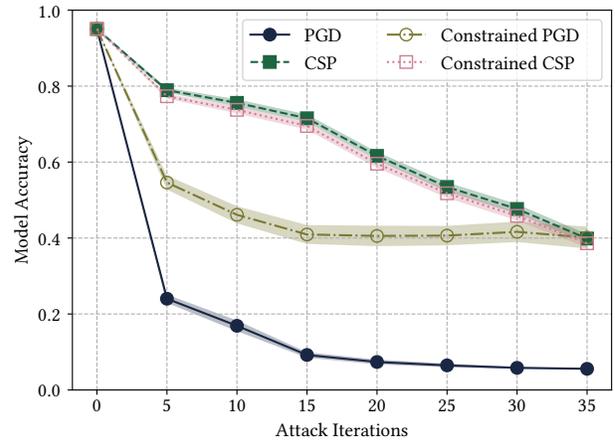}}
        \caption{\ph{} --- Model Accuracy}\label{fig:results:phishing:accuracy}
    \end{subfigure}
    \caption{Crafting Adversarial Examples with Domain Constraints---Rate of
    invalid samples and model accuracy as a function of attack iterations
    (iterations add \(\pm 0.01\) to continuous features and \SI{\pm{}1} to
    categorical features). \csp{} and \pgd{} refer to the attacks
    pre-projection, while \texttt{Constrained} variants demonstrate results
    post-projection with \dpll{}. Shaded regions represent \SI{95}{\percent}
    C.I.\vspace{-3mm}}\label{fig:results}
\end{figure*}

\shortsection{Measuring the Efficacy of Constraints} Figure~\ref{fig:results}
illustrates the rate of invalid samples (that is, the number of constraint
non-compliant adversarial examples over the total number crafted) and model
accuracy as a function of the number of iterations used to craft adversarial
examples.

The rate of invalid samples (Figures~\ref{fig:results:nsl:rate}
and~\ref{fig:results:phishing:rate}) answers our first evaluation question:
\textit{do the known crafting algorithms violate domain constraints?} Across
both the \nsl{} and \ph{}, we observe that \textit{at least} \SI{60}{\percent}
of all adversarial examples produced by \pgd{} violate the domain constraints
when the number of iterations exceeds \SI{10}{}. In cases where domain
constraints are violated, the adversarial examples produced are not realizable.

For our second evaluation question: \textit{Do constraints add robustness?} The
results suggest that domain constraints add robustness against adversarial
examples.  For example, we observe that even though \dpll{} was largely able to
successfully project the adversarial examples produced by \pgd{} when the
number of iterations was small, many of the resultant constraint-compliant
adversarial (\ie{} \textsc{Constrained-PGD}) examples were correctly classified
by the model. Notably, we observe how \SI{34}{\percent} of model accuracy is
restored in the \ph{} dataset (Figure~\ref{fig:results:phishing:accuracy}) once
invalid examples produced by \pgd{} are projected back onto a
constraint-compliant space. On the \nsl{} dataset, running \pgd{} with many
iterations produced additional examples that could not be projected into the
constraint-compliant space, increasing the accuracy of the model.

Finally, we examine the applicability of \csp{} to crafting valid adversarial
examples. The conservative nature of the \csp{} lends itself to producing
adversarial examples that readily comply with domain constraints: at 10
iterations, only \SI{10}{\percent} of the examples produced by the \csp{}
violated domain constraints in the worst case (compared to nearly
\SI{60}{\percent} for \pgd{}). Additionally, while many examples produced by
\pgd{} cannot be projected onto the constraint-compliant space (about
\SI{40}{\percent} for the \ph{} dataset), examples produced by \csp{} were
successfully projected nearly \SI{100}{\percent} of the time. This suggests
that saliency-based algorithms, as well as algorithms targeting an \lp{0} norm,
may be more readily applicable to constrained domains than gradient-based
algorithms or those targeting \lp{\infty}. This is consistent with the
structure of constraints and budget used by \dpll{}; \pgd{} causes larger
perturbations over \lp{0}, which will likely violate more constraint clauses
and frustrate projection.

\shortsection{Takeaway} From our investigation, we highlight key takeaways: (A)
\textit{Crafting constraint-compliant adversarial examples is a necessarily
different process than traditional crafting approaches}. Up to
\SI{82}{\percent} of adversarial examples produced by \pgd{} violated domain
constraints. (B) \textit{Constraints add robustness}. In the worst case,
\SI{34}{\percent} of model accuracy was restored after projecting adversarial
examples onto the learned constraint theory.

The results demonstrate that crafting adversarial examples in constrained
domains is a \textit{necessarily} different process than those of unconstrained
domains. Domain constraints have a tangible impact on the underlying threat
surface as many of the threats produced by known crafting algorithms are not
\textit{realizable}. Perhaps most importantly, the relationships between
features serve as a form of robustness to the known crafting algorithms.

\subsection{Scalability}\label{evaluation:scalability}

We next consider the scalability of our approach. Recall that Valiant's
Algorithm (Algorithm~\ref{alg:valiant}) checks each constraint against each
observation, returning only constraints that certify all observations.
Valiant's algorithm, therefore, has time complexity \(\bigo{}(|E|\cdot|T|)\)
(Note that \(|T|=\bigo{}(|C|)\)).  For clauses of cardinality \(k=1\),
\(|C|=\prod_{X_i\in X} |X_i|\), and the algorithm takes time
\(\bigo{}(\prod{}|X_i|)\). Thus the combined runtime is
\(\bigo{}(\prod{}|X_i|+|E|\cdot|T|)=\bigo{}(|E|\cdot \prod{}|X_i|)\).

\begin{table}
    \begin{tabular}{lcccc}
        \toprule
        Datasets & \(|E|\) & \(\prod{}|X_i|\) & \(d\) & \(\frac{d}{|E|\cdot \prod{}|X_i|}\) \\
        \midrule
        \nsl{}    & \SI{1.5E5}{} & \SI{1.0E4}{} & \(\SI{180}{s}\)  & \(\SI{1.2E-07}{s}\) \\
        \ph{}     & \SI{1.0E4}{} & \SI{8.4E3}{} & \(\SI{7}{s}\)    & \(\SI{8.3E-08}{s}\) \\
        \ddos{}   & \SI{2.5E6}{} & \SI{1.3E4}{} & \(\SI{4560}{s}\) & \(\SI{1.4E-07}{s}\) \\
        \drebin{}~\cite{arp_drebin_2014} (\textit{est.}) & \SI{1.2E5}{}  & \SI{5.6E5}{} & \(\SI{8046}{s}\)  & --- \\
        \aninea{}~\cite{kohavi_scaling_1996} (\textit{est.}) & \SI{4.6E4}{}  & \SI{7.6E6}{} & \(\SI{39500}{s}\) & --- \\
        \midrule
        \multicolumn{4}{c}{Mean} & $\SI{1.1E-07}{s}$ \\
        \bottomrule
    \end{tabular}
    \caption{Measured and estimated time to learn constraints\vspace{-5mm}}\label{tab:scalability}
\end{table}

\begin{figure}
    \centering
    \resizebox{\columnwidth}{!}{\input{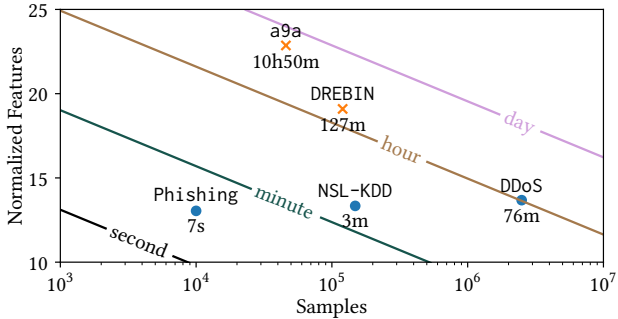}}
    \caption{Scalability of constraint learning. Contours show time to learn
    constraints based on samples and features. Feature counts have been
    normalized ($\log_2 \prod{}|X_i|$) to approximate equivalent binary
    features. For comparison we superimpose actual and estimated runtimes of
    various datasets}\label{fig:dst_plot}
\end{figure}

Next, we explore how the approach scales with real datasets.
Table~\ref{tab:scalability} shows, for each dataset, the number of unique
samples \(|E|\), the size of the power set and the total run time to execute
the learning process.  In addition to the \nsl{} and \ph{} datasets and to
observe performance under large \(|E|\), we measure performance on the large
\ddos{} dataset~\cite{sharafaldin_developing_2019} which contains 2.5 million
unique samples, after applying feature reduction techniques, as shown in
~\cite{li_detection_nodate}, from a complex network environment.  We also
provide estimated performance on two additional datasets after feature
reduction techniques inspired by literature.

Figure~\ref{fig:dst_plot} visualizes the performance of the learning process
over measured and estimated datasets.  Here we show  (1) the number of samples
in the training data, and (2) the number of features. To allow comparison
between datasets with differing classes per feature, we use a normalized
feature count, which is the equivalent number of binary features which would
yield the same \(\prod |X_i|\). We superimpose actual and estimated performance
numbers for comparison. From this, it can be seen that compute time scales
linearly with the number of samples, but exponentially with the number of
features. Because the number of samples in a training dataset should generally
increase exponentially with the number of features (\ie{} the curse of
dimensionality~\cite{bellman_adaptive_1961}: the number of features should be
\(\bigo{}(\log|E|)\)), the complexity of learning constraints on well-formed
datasets can be roughly modeled as \(\bigo{}(|E|^2)\). As previously noted,
feature reduction techniques can further increase the tractability of
high-dimensionality datasets without significantly lowering model accuracy.
Additionally, further optimizations in the constraint learning routine might
greatly improve throughput.

As a final note, recall our use of \optics{} for clustering continuous
variables; we observed significant performance bottlenecks here. \optics{}
computes pairwise distances between points that are within a parameterized
\(\epsilon\) neighborhood distance. For completeness, we set \(\epsilon\) to
\(\infty\), which yields a runtime of \(\bigo{}(n^2)\). Depending on the number
of samples, this can take \textit{days} to determine the clusters alone (for
comparison, once the clusters are identified, constraint learning can be done
on the order of minutes). Supplementary experiments did show that \optics{}
could approximate clusters fairly well with \SI{1}{\percent} randomly sampled
subsets of the training data (\eg{} for the \nsl{}, out of the 18 clusters
identified across features with the full training set, 16 clusters were
consistently correctly identified with \SI{1}{\percent} randomly sampled
subsets of the training set).

\section{Discussion}\label{discussion}

\shortsection{Projecting vs. Enforcing Constraints} One of the core components
of the \csp{} is using \dpll{} to project adversarial examples onto the
constraint compliant space, characterized by \(T\). In preliminary experiments,
we enforced \(T\) throughout the crafting process, that is, at any arbitrary
iteration \(r\), \(e^{r}\) was always a realizable adversarial example.
However, we observed that, in some cases, the domain constraints would create
``archipelagos'' around inputs; any perturbations to an input were bound to
small regions of the input space.  Unconstrained domains, however, have input
spaces that are more akin to supercontinents. This suggests that, for many
domains deploying machine learning, the threat surface exposed by vanilla
applications of adversarial machine learning is an overestimation (sometimes a
large one) of the true, practical threat surface.

\shortsection{Robustness \(\cup\)
Constraints}~\cite{raghunathan_certified_2018, wong_provable_2018} show two
techniques for providing formal guarantees on model robustness (\ie{}
homogeneous predictions in a \lp{\infty}-norm ball). While these approaches
have (at present) limitations that discourage practical deployment, they
convincingly suggest that a model robust to adversaries is attainable. We note
that there were no algorithms applied to our models to ``secure'' them, yet we
observed tangible gains in model robustness from domain constraints.  This
suggests that the pairing of some form of model robustness with domain
constraints may produce models that are highly challenging for an adversary to
exploit. We present this as an opportunity for future work.

\shortsection{Addressing Concept Drift}\label{discussion:concept_drift} Many
applications of machine learning involve \textit{non-stationary phenomena}; as
the underlying phenomena changes over time, so does the space of observations
that comply with the domain constraints.  As a natural consequence of learning
constraints from data, learning relevant constraint theories may necessitate:
(1) identifying when concept drift has occurred, and (2) rectifying its effects
on the learned constraint theory. We describe below several experiments that
characterize the effects of concept drift on constraint learning.

Identifying concept drift is an unavoidable problem in machine learning, and
there are approaches that can dynamically recognize concept drift and react to
mitigate its effects (such as moving windows or detection
thresholds)~\cite{webb_characterizing_2016, widmer_learning_1996,
gama_learning_2004}. As a measurement of detecting concept drift, we
investigated whether constraint theories learned exclusively from the training
set would reject observations from the test set.  Intuitively, this exercise is
useful for two reasons: (1) it informs us if our approach overfits to a set of
collected data (\ie{} does Valiant's algorithm \textit{generalize}?), and (2)
emulates what practical deployments of constraint learning would observe, given
that datasets with a dedicated test set include new observations to approximate
excepted performance when deployed.

\begin{table}
    \begin{tabular}{lccc}
        \toprule
        Dataset & \# Clauses & \# Violations & \% Violations \\
        \midrule
        \nsl{}  & \(5,874\) & 401 & \SI{1.7}{\percent}  \\
        \ph{}   & \(2,143\)  & 14   & \SI{0.45}{\percent} \\
        \bottomrule
    \end{tabular}
    \caption{Constraint Violations from Test Set Observations\vspace{-5mm}}\label{tab:cd}
\end{table}

Our results, shown in Table~\ref{tab:cd} show a stark contrast in the violation
rate between inputs from the test set and the adversarial examples crafted in
our evaluation. These results confirm that: (1) Valiant's algorithm does learn
constraint theories that generalize well on unseen data, and, more importantly,
(2) constraint violations can serve as an indicator of concept drift. For
example, in a practical setting, network operators could use constraint
violations as a ``filter'' for traffic flows that deserve attention and if the
violation rate were to exceed some threshold, an indicator of concept drift
through changes in the underlying traffic patterns.

Once concept drift has been identified, \textit{rectifying} its effects can be
challenging. Under the framework evaluated in this work, there are two facts on
the effects of concept drift to a learned constraint theory: (1) new
observations remove clauses that encode dated constraints, and (2) old
observations could have removed clauses that encode constraints irrelevant at
that time, but could be applicable at present (for example, a service could
drop support for a particular protocol in future versions). In our setting,
addressing dated constraints is straightforward: new observations could be used
to remove the irrelevant constraints in a linear pass. However, re-adding
constraints that were irrelevant in the past requires a more nuanced approach.

The naive approach would simply be to re-generate the universe of constraints
and repeat the learning process in its entirety. However, this can be
time-prohibitive for some domains, and so we are keen to use an approach were
complete re-learning is not necessary. To this end, we are inspired by specific
approach used in~\cite{widmer_learning_1996} to mitigate the effects of concept
drift. Specifically,~\cite{widmer_learning_1996} stores concept descriptions
and reuses them when a particular context appears. Thematically similar, we
could first record the clauses removed for all observations in a training set.
Then, if an observation is considered to be a dated representation of the
domain (we could leverage domain expertise to identify such observations), we
could simply re-add the removed clauses (insofar as no other observations would
also remove those clauses). In this way, we can trade off repeating the
learning process versus maintaining a record for the clauses removed for all
samples.

\shortsection{The Quality of Learned Constraints} Ultimately the goal of this
work is to develop constraints that reflect the true limitations of the domain.
Historically, this has been the purview of domain experts. For example, Stolfo
et al. developed a comprehensive constraints for network
traffic~\cite{stolfo_cost-based_2000}. Below, we show that by example the
constraint theory learns constraints identified by humans (including those by
Stolfo et al. and a number developed from our own experience with IP network
protocols). We are exploring a more exhaustive analysis that systematically
compares the specifications of IP protocols to the learned constraints.

To compare learned and human cultivated constraints, we formulate queries in
the form of inputs that demonstrate constraints we would expect the constraint
theory to learn. From our queries, we verified both obvious and subtle
constraints, namely (1) if a TCP flow was terminated with \texttt{REJ} flag
(\ie{} through the source sending an initial \texttt{SYN} packet with the
\texttt{RST} bit set), then the number of bytes sent in the flow must be \(0\)
as, for the \nsl{}, bytes measured in a flow was done post-handshake), and (2)
\texttt{SYN} packets that have the same IP and port numbers for the source and
destination fields, flagged as ``land'' in the \nsl{}, are never responded to.
While the two above serve as examples of constraints from the \texttt{TCP/IP}
protocol and domain experts respectively, the constraint theory also learned
some attack-specific constraints, such as flows that had errors at some stage
of the TCP handshake towards a specific service were distributed across
destinations. After analyzing the kinds of flows that exhibited this property
in the dataset, we observed that this was almost exclusively associated with
probe attacks---this agrees with our understanding of probe attacks in that an
adversary seeks to collect information of the services available in a network
across destination hosts through a ``heartbeat'' mechanism, such as initiating
connections to observe any form of response.

\section{Related Work}\label{related}

\shortsection{Learning in the Presence of Adversaries} The origins of
adversarial machine learning are not explicitly known and are often
disputed~\cite{biggio_wild_2018}. From our perspective, exploring the degree to
which adversaries can influence learning algorithms begins in \si{1993} with
Kearns et al. who formalize a worst-case data poisoning attack for any learning
algorithm~\cite{kearns_learning_1993}. In
\si{1997},~\cite{uther_adversarial_1997} explores the efficacy of reinforcement
algorithms in adversarial environments, with \textsc{Minimax} tables driving
agent (and adversary) decisions.

\shortsection{The Rise of Deep Learning} With the rise in popularity in deep
learning, adversarial scenarios were
revisited~\cite{goodfellow_explaining_2014, papernot_limitations_2016,
szegedy_intriguing_2013, carlini_towards_2017}. Many early works explored
white-box, inference-time attacks via gradient-based algorithms. Shortly after,
adversarial methods were translated from conceptual to practical, as works
demonstrated how to produce adversarial examples in physical spaces, using
stickers, glasses, and graffiti~\cite{kurakin_adversarial_2016,
brown_adversarial_2017, eykholt_robust_2018, sharif_general_2019}.
Subsequently, adversarial machine learning was no longer exclusive to academia;
it began to enter popular culture with discussions of fooling AI in magazine
articles~\cite{kobie_cripple_2018, mok_google_2018, gershgorn_fooling_2019},
demonstrating ``DeepFakes'' on television~\cite{leetaru_what_2019}, and even
displaying adversarial examples in museums~\cite{hitti_science_2019}.

In \si{2017} and \si{2018}, there was a burst of adversarial machine learning
research; transfer attacks (\ie{} grey-box)~\cite{dong_evading_2019,
kurakin_adversarial_2017}, black-box attacks on machine-learning-as-a-service
platforms~\cite{brendel_decision-based_2017, papernot_practical_2017,
ilyas_black-box_2018, papernot_transferability_2016}, attacks by altering a
single feature~\cite{su_one_2019}, data poisoning attacks (\ie{} at training
time)~\cite{alfeld_data_2016, chen_targeted_2017}, adversarial example
detectors~\cite{xu_feature_2017, feinman_detecting_2017,
carlini_adversarial_2017, grosse_statistical_2017}, adversarially robust models
through adversarial training~\cite{madry_towards_2017}, linear
programming~\cite{wong_provable_2018}, and semidefinite
relaxations~\cite{raghunathan_certified_2018, steinhardt_certified_2017}, among
many other works. Seemingly every corner of machine learning involving some
form of an adversary was explored.

\shortsection{Images and Beyond} As we motivated in Section~\ref{introduction},
the majority of applications in adversarial machine learning have been in
images. Recently, we have started to see applications in security domains,
including malware~\cite{kolosnjaji_adversarial_2018, grosse_adversarial_2017,
anderson_evading_nodate}, and network intrusion
detection~\cite{yang_adversarial_2018, lin_idsgan_2018,
rigaki_adversarial_2017, sheatsley_adversarial_2020}. These works all describe
similar motivations: domains concerned with adversarial machine learning will
likely not be exclusive to images. As canonical representatives of security,
malicious software and network traffic are relevant phenomena to study.
However, there are commonalities among the works that limit the applicability
of the findings to practical deployments.

When perturbing inputs, these works exploit \textit{domain ``safe-spaces''}.
For malware applications, the authors acknowledge that perturbing malware
directly is incredibly challenging (without breaking it or removing its
malicious purpose), therefore, perturbations are limited to either appending
bytes at the end of the binary~\cite{kolosnjaji_adversarial_2018} or only
adding permissions to the list of required permissions for an application, in
the context of Android malware~\cite{grosse_adversarial_2017}. These are two
examples of ``safe-space'' perturbations: such manipulations guarantee that
malicious behavior is preserved (and that the malware is still functional) by
identifying regions that explicitly \textit{avoid} domain constraints. This
effectively reduces to manipulating image-like inputs, in that perturbations
can be applied arbitrarily and independently.

For the works in network intrusion detection, some either ignore domain
constraints~\cite{rigaki_adversarial_2017, yang_adversarial_2018}, or rely on
domain expertise to identify what can and what cannot be
perturbed~\cite{lin_idsgan_2018, sheatsley_adversarial_2020}.
Specifically,~\cite{lin_idsgan_2018} argues that insofar as the identified
features are not perturbed, then the malicious behavior is preserved. Not
unlike the malware scenarios, this models adversaries as being able to perturb
arbitrarily and independently (just with a reduction to the allowable
perturbation space, much like the one-pixel attack in~\cite{su_one_2019}).
However,~\cite{sheatsley_adversarial_2020} provides a method where all features
can be perturbed with a subroutine to enforce the domain constraints. The
approach suffers from relying on expertise to identify the constraints
correctly and is largely formulated for network intrusion detection systems.

While in this work we use images as a motivating example of an unconstrained
domain, domain constraints can exist in images, depending on the subject of the
image. Specifically, Chandrasekaran et al.\ identify domain constraints in
images with domain expertise (as well as a data-driven approach via embeddings
at intermediary layers of the model)~\cite{chandrasekaran_rearchitecting_2019}.
They perform a complementary observation that adversarial crafting algorithms
violate domain constraints, and, when domain constraints are enforced, model
robustness is improved.

\shortsection[?]{Where To} While we are moving closer to accurate threat models
for diverse domains, we may ask, ``What if such safe-spaces do not exist? What
if the adversary is \textit{required} to perturb in regions that may have
effects on other features? Can such adversarial examples be realized?'' We
argue that these are fundamental questions for any domain that is keen to
deploy machine learning.

\section{Conclusion}\label{conclusion}

This paper explored adversarial examples with \textit{domain constraints}:
relationships between features that encode the rules or structures of the
underlying phenomena. We develop algorithms to learn constraint theories for
given data distributions and integrate domain constraints into adversarial
crafting processes. By representing domain constraints as logic clauses, we
design a data-driven approach to learn the domain constraints across network
intrusion detection and phishing datasets. We find that: (1) crafting
adversarial examples in constrained domains is a \textit{necessarily} different
process than unconstrained domains; up to \SI{82}{\percent} of adversarial
examples produced by \pgd{} violated domain constraints, and (2) constrained
domains are inherently more robust against adversarial examples; in one domain,
\SI{34}{\percent} of model accuracy was restored after projecting adversarial
examples onto the learned constraint theory. These findings suggest that the
exploitable threat surface of models in constrained domains is likely narrower
than previously understood.

\begin{acks}

    The authors would like to warmly thank the reviewers for their insightful
    feedback and Quinn Burke for his feedback on versions of this paper. This
    research was sponsored by the Combat Capabilities Development Command Army
    Research Laboratory and was accomplished under Cooperative Agreement Number
    W911NF-13-2-0045 (ARL Cyber Security CRA). The views and conclusions
    contained in this document are those of the authors and should not be
    interpreted as representing the official policies, either expressed or
    implied, of the Combat Capabilities Development Command Army Research
    Laboratory or the U.S. Government. The U.S. Government is authorized to
    reproduce and distribute reprints for Government purposes not withstanding
    any copyright notation here on. This material is based upon work supported
    by the National Science Foundation under Grant No. CNS-1805310. Any
    opinions, findings, and conclusions or recommendations expressed in this
    material are those of the author(s) and do not necessarily reflect the
    views of the National Science Foundation.

\end{acks}

\bibliographystyle{plain}
\balance{}
\bibliography{references}
\clearpage\appendix\section{The Worst-Case Adversary}\label{appendix-a}

We call a constraint theory \(T\) \textit{strict} when \(T\) \textit{only}
certifies the observations \(E\) from which \(T\) was learned; any other
observation \(e^*\) that is not a member of \(E\) is rejected. Secondly, we
call \(T\) \textit{general} when it certifies a \textit{maximum} number of
observations (while containing a non-zero number of clauses).

We will now show that, in the setting described in this paper, the \textit{most
strict} constraint theories are those whose clauses have literals of
cardinality \(n-1\) (recall, literals in our setting are sets), where \(n\)
describes the number of unique values observed for a given attribute in the set
of collected observations \(E\). Such constraint theories describe the
\textit{best-case} adversary, as the adversary must produce adversarial
examples that are precise copies of collected observations \(E\).

Conversely, the \textit{most general} constraint theories are those whose
clauses have literals of cardinality \(1\). These constraint theories
characterize the \textit{worst-case} adversary, as such constraint theories
certify the \textit{maximum} number of observations among all constraint
theories learned by considering literals with cardinalities from \(1\) to
\(n-1\). Let \(\psi_k\) be the set of observations rejected by constraint
theories $T_k$ whose clauses contain literals with cardinality \(k\). We will
now show: \(\psi_1\subseteq\cdots\subseteq\psi_k\subseteq\cdots\psi_{n-1}\)

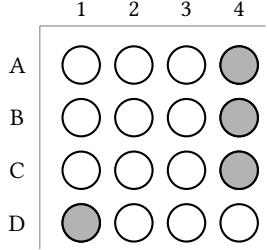
\begin{figure}[!h]
    \begin{tikzpicture}
        \coordinate (Origin)   at (0,0);
        \coordinate (XAxisMin) at (0.15,3.32);
        \coordinate (XAxisMax) at (3.15,3.32);
        \coordinate (YAxisMin) at (0.15,0.35);
        \coordinate (YAxisMax) at (0.15,3.32);
        \draw [thin, gray] (XAxisMin) -- (XAxisMax);
        \draw [thin, gray] (YAxisMin) -- (YAxisMax);

        \foreach \p in {1,2,3,4}{%
            \node[fill=none] at (0.7*\p,3.55) {\p};}
        \foreach \p\v in {1/D,2/C,3/B,4/A}{%
            \node[fill=none] at (-0.15,0.7*\p) {\v};}

        \foreach \x in {1,2,3,4}{%
            \foreach \y in {1,2,3,4}{%
                \node[thick,draw,circle,inner sep=5pt,fill=white!60,
                name=circle-\x-\y] at (0.7*\x,0.7*\y) {};}}

        \node[thick,draw,circle,inner sep=5pt,fill=gray!60] at (0.7*1,0.7*1){};
        \foreach \y in {2,3,4}{%
            \node[thick,draw,circle,inner sep=5pt,fill=gray!60] at (0.7*4,0.7*\y){};}
    \end{tikzpicture}
    \caption{Domain and collected observations \(E\)---In the considered
    domain, there are two features \(x_1\) and \(x_2\) whose observations can
    take values \(\set{A,B,C,D}\) and \(\set{1,2,3,4}\), respectively. The grey
    circles represent the set of collected observations
    \(E\).}\label{fig:proof_e}
\end{figure}

\shortsection{An Illustrative Domain} Consider the visualization in
Figure~\ref{fig:proof_e} of some set of collected observations \(E\) (shaded in
gray).

In this domain, there are two features: \(x_1\), which can take values
\(\set{A,B,C,D}\), and \(x_2\), which can take values \(\set{1,2,3,4}\). From
\(E\), we observe that if some observation \(e\) has value \(x_1=D\), then
\(x_2=1\), otherwise if \(x_1\neq D\), then \(x_2=4\).

Next, we will consider the output of Algorithm~\ref{alg:gc}, generating the
space of possible of constraints, for literals of cardinality \(k=1,2\), and
\(n-1=3\). For clarity, a single clause will be written as \((\set{\alpha}\lor
\set{\beta})\), where \(\alpha\) represents at least one and at most three
elements from \(x_1\) (\ie{} \(\set{A,B,C,D}\)), and \(\beta\) represents at
least one and at most three elements from \(x_2\) (\ie{} \(\set{1,2,3,4}\)).

\shortsection{\(\boldsymbol{k=1}\)} Let us consider the space of possible
constraints for clauses generated with literals of cardinality \(k=1\):

\begin{align*}
    (\set{A}\lor\set{1})\land(\set{A}\lor\set{2})&\land(\set{A}\lor\set{3})\land(\set{A}\lor\set{4})\land\\
    (\set{B}\lor\set{1})\land(\set{B}\lor\set{2})&\land(\set{B}\lor\set{3})\land(\set{B}\lor\set{4})\land\\
    (\set{C}\lor\set{1})\land(\set{C}\lor\set{2})&\land(\set{C}\lor\set{3})\land(\set{C}\lor\set{4})\land\\
    (\set{D}\lor\set{1})\land(\set{D}\lor\set{2})&\land(\set{D}\lor\set{3})\land(\set{D}\lor\set{4})
\end{align*}

After applying Valiant's algorithm to these constraints with our set of
collected observations \(E\), the resultant learned constraint theory is then:

\noindent\begin{equation*}
    T=(\set{D}\lor\set{4})
\end{equation*}

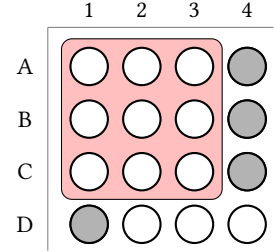
\begin{figure}[!h]
    \begin{tikzpicture}
        \coordinate (Origin)   at (0,0);
        \coordinate (XAxisMin) at (0.15,3.32);
        \coordinate (XAxisMax) at (3.15,3.32);
        \coordinate (YAxisMin) at (0.15,0.35);
        \coordinate (YAxisMax) at (0.15,3.32);
        \draw [thin, gray] (XAxisMin) -- (XAxisMax);
        \draw [thin, gray] (YAxisMin) -- (YAxisMax);
        \foreach \p in {1,2,3,4}{%
            \node[fill=none] at (0.7*\p,3.55) {\p};}
        \foreach \p\v in {1/D,2/C,3/B,4/A}{%
            \node[fill=none] at (-0.15,0.7*\p) {\v};}

        \foreach \x in {1,2,3,4}{%
            \foreach \y in {1,2,3,4}{%
                \node[thick,draw,circle,inner sep=5pt,fill=white!60,
                name=circle-\x-\y] at (0.7*\x,0.7*\y) {};}}

        \foreach \y in {2,3,4}{%
            \node[thick,draw,circle,inner sep=5pt,fill=gray!60,
            name=circle-\y] at (0.7*4,0.7*\y) {};}
        \node[thick,draw,circle,inner sep=5pt,fill=gray!60] at (0.7*1,0.7*1){};

        \begin{pgfonlayer}{bg}
            \node[draw, fill=red!25, rectangle, rounded corners, fit={(circle-1-2)
            (circle-3-4)}] {};
        \end{pgfonlayer}
    \end{tikzpicture}
    \caption{Reject space for \(k=1\)---For literals whose cardinality \(k\)
    is \(1\), the learned constraint theory \(T=(\set{D}\lor\set{4})\) rejects
    any observation from the red box.}\label{fig:proof_k1}
\end{figure}

As we described above, our worst-case adversary is one who crafts adversarial
examples for a constraint theory that is \textit{most general}, among all
possible learned constraint theories in our setting. Consider the learned
constraint theory \(T=(\set{D}\lor\set{4})\); this constraint theory will
certify any observation \(e\) whose values are either \((D,\cdot)\) or
\((\cdot,4)\) (where \(\cdot\) denotes any value from the domain of the
respective attribute). From Figure~\ref{fig:proof_e}, we can see that, out of
the \(16\) possible instances in our exemplar domain, seven are accepted and
nine are rejected, that is, \(|\psi_1|=9\). Now, we draw a ``reject'' box in
red, respectively, shown in Figure~\ref{fig:proof_k1}.

Reject boxes can be derived by inverting each individual clause and applying De
Morgan's law. For example, the reject box associated with clause
\((\set{D}\lor\set{4})\) would be
\(\overline{(\set{D}\lor\set{4})}=(\overline{\set{D}}\land\overline{\set{4}})=(\set{A,B,C}\land\set{1,2,3})\).
We can interpret this reject box as: \(T\) \textit{will reject any observation
with} \(x_1\in\set{A,B,C}\) and \(x_2\in\set{1,2,3}\); \textit{otherwise,}
\(T\) \textit{will certify it.}

As another important remark, beyond the observations in \(E\) used to learn
\(T\), \(T\) \textit{will also certify the unseen observations} \((D,2),
(D,3)\) \textit{and} \((D,4)\). This is an example of the
\textit{generalization} provided by encoding constraints of cardinality
\(k=1\).

\shortsection{\(\boldsymbol{k=2}\)} We now continue with learning a constraint
theory with cardinality \(k=2\). Again, consider the space of possible
constraints whose literals have cardinalities \(k=2\):

\begin{align*}
    (\set{A,B}\lor\set{1,2})\land(\set{A,B}&\lor\set{1,3})\land(\set{A,B}\lor\set{1,4})\land\\
    (\set{A,B}\lor\set{2,3})\land(\set{A,B}&\lor\set{2,4})\land(\set{A,B}\lor\set{3,4})\land\\
    (\set{A,C}\lor\set{1,2})\land(\set{A,C}&\lor\set{1,3})\land(\set{A,C}\lor\set{1,4})\land\\
    (\set{A,C}\lor\set{2,3})\land(\set{A,C}&\lor\set{2,4})\land(\set{A,C}\lor\set{3,4})\land\\
    (\set{A,D}\lor\set{1,2})\land(\set{A,D}&\lor\set{1,3})\land(\set{A,D}\lor\set{1,4})\land\\
    (\set{A,D}\lor\set{2,3})\land(\set{A,D}&\lor\set{2,4})\land(\set{A,D}\lor\set{3,4})\land\\
    (\set{B,C}\lor\set{1,2})\land(\set{B,C}&\lor\set{1,3})\land(\set{B,C}\lor\set{1,4})\land\\
    (\set{B,C}\lor\set{2,3})\land(\set{B,C}&\lor\set{2,4})\land(\set{B,C}\lor\set{3,4})\land\\
    (\set{B,D}\lor\set{1,2})\land(\set{B,D}&\lor\set{1,3})\land(\set{B,D}\lor\set{1,4})\land\\
    (\set{B,D}\lor\set{2,3})\land(\set{B,D}&\lor\set{2,4})\land(\set{B,D}\lor\set{3,4})\land\\
    (\set{C,D}\lor\set{1,2})\land(\set{C,D}&\lor\set{1,3})\land(\set{C,D}\lor\set{1,4})\land\\
    (\set{C,D}\lor\set{2,3})\land(\set{C,D}&\lor\set{2,4})\land(\set{C,D}\lor\set{3,4})
\end{align*}

After applying Valiant's algorithm to this batch of clauses with \(E\), the
learned constraint theory is then:

\noindent\begin{align*}
    T=(\set{A,B}\lor\set{1,4})\land(\set{A,C}&\lor\set{1,4})\land(\set{A,D}\lor\set{1,4})\land\\
      (\set{A,D}\lor\set{2,4})\land(\set{A,D}&\lor\set{3,4})\land(\set{B,C}\lor\set{1,4})\land\\
      (\set{B,D}\lor\set{1,4})\land(\set{B,D}&\lor\set{2,4})\land(\set{B,D}\lor\set{3,4})\land\\
      (\set{C,D}\lor\set{1,4})\land(\set{C,D}&\lor\set{2,4})\land(\set{C,D}\lor\set{3,4})
\end{align*}

Here, we observe that the union of the reject boxes produced by clauses that
contain \(D\) \textit{is exactly identical to the reject box produced by
clause} \((\set{D}\lor\set{4})\). Any observation will only satisfy those
clauses in \(T\) if the observation has \(x_1=D\) or \(x_2=4\). However, we do
note that there are some unique clauses that produce reject boxes that are not
a direct subset of the reject box produced by clause \((\set{D}\lor\set{4})\).
Namely, \((\set{A,B}\lor\set{1,4})\), \((\set{A,C}\lor\set{1,4})\), and
\((\set{B,C}\lor\set{1,4})\). Of these unique clauses, without loss of
generality, consider the reject box produced by \((\set{A,B}\lor\set{1,4})\),
that is: \((\set{C,D}\land\set{2,3})\).

\begin{figure}[!h]
    \begin{tikzpicture}
        \coordinate (Origin)   at (0,0);
        \coordinate (XAxisMin) at (0.15,3.32);
        \coordinate (XAxisMax) at (3.15,3.32);
        \coordinate (YAxisMin) at (0.15,0.35);
        \coordinate (YAxisMax) at (0.15,3.32);
        \draw [thin, gray] (XAxisMin) -- (XAxisMax);
        \draw [thin, gray] (YAxisMin) -- (YAxisMax);
        \foreach \p in {1,2,3,4}{%
            \node[fill=none] at (0.7*\p,3.55) {\p};}
        \foreach \p\v in {1/D,2/C,3/B,4/A}{%
            \node[fill=none] at (-0.15,0.7*\p) {\v};}

        \foreach \x in {1,2,3,4}{%
            \foreach \y in {1,2,3,4}{%
                \node[thick,draw,circle,inner sep=5pt,fill=white!60,
                name=circle-\x-\y] at (0.7*\x,0.7*\y) {};}}

        \foreach \y in {2,3,4}{%
            \node[thick,draw,circle,inner sep=5pt,fill=gray!60,
            name=circle-\y] at (0.7*4,0.7*\y) {};}
        \node[thick,draw,circle,inner sep=5pt,fill=gray!60] at (0.7*1,0.7*1){};

        \begin{pgfonlayer}{bg}
            \node[draw, fill=red!25, rectangle, rounded corners, fit={(circle-1-2)
            (circle-3-4)}] {};
        \end{pgfonlayer}
        \begin{pgfonlayer}{bg}
            \node[draw, fill=orange!25, rectangle, rounded corners, fit={(circle-2-1)
            (circle-3-2)}] {};
        \end{pgfonlayer}
    \end{tikzpicture}
    \caption{Reject space for \(k=2\)---In addition to the observations
    rejected for \(k=1\) (shown in red), observations \((D,2)\) and \((D,3)\)
    are also rejected when \(k=2\) (shown in orange).}\label{fig:proof_k2}
\end{figure}
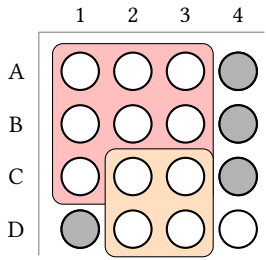

Now, we draw a reject box for this clause in orange and the union of the reject
boxes for clauses containing \(D\) in red, shown in Figure~\ref{fig:proof_k2}.
In this setting, we can see that the learned constraint theory is \textit{less
general}: \(T\) certifies five observations and rejects eleven (\ie{}
\(|\psi_2|=11\)). Thus far, we have shown \(|\psi_1|<|\psi_2|\).

From this, we make two important observations: (1) the size of the reject boxes
are inversely proportional to the cardinality of the literals (\ie{} the
constraints are becoming \textit{more granular} and \textit{less general}), and
(2) in addition to the observations rejected when \(k=1\), observations
\((D,2)\) and \((D,3)\) are now also rejected by \(T\). In the context of
generalization, beyond the observations used to learn \(T\), only the unseen
observation \((D,4)\) will be certified by \(T\) when \(k=2\).

\shortsection{\(\boldsymbol{k=n-1=3}\)} We now finish with learning a
constraint theory with cardinality \(k=3\) (note that, for this domain,
\(n=4\), and thus, any clause with literals of cardinality four will be
axiomatically satisfied, and so the maximum cardinality we consider is
\(n-1=3\)). We compute the last batch of the space of possible of constraints
for \(k=3\):

\begin{align*}
    (\set{A,B,C}\lor\set{1,2,3})\land(\set{A,B,C}&\lor\set{1,2,4})\land(\set{A,B,C}\lor\set{1,3,4})\land\\
    (\set{A,B,C}\lor\set{2,3,4})\land(\set{A,B,D}&\lor\set{1,2,3})\land(\set{A,B,D}\lor\set{1,2,4})\land\\
    (\set{A,B,D}\lor\set{1,3,4})\land(\set{A,B,D}&\lor\set{2,3,4})\land(\set{A,C,D}\lor\set{1,2,3})\land\\
    (\set{A,C,D}\lor\set{1,2,4})\land(\set{A,C,D}&\lor\set{1,3,4})\land(\set{A,C,D}\lor\set{2,3,4})\land\\
    (\set{B,C,D}\lor\set{1,2,3})\land(\set{B,C,D}&\lor\set{1,2,4})\land(\set{B,C,D}\lor\set{1,3,4})\land\\
    (\set{B,C,D}\lor\set{2,3,4})
\end{align*}

Here, it is worth noting an important observation: \textit{each of these
clauses will fail to be satisfied by one and only one observation} (\ie{} the
size of the reject boxes for each clause at \(k=3\) is one-by-one). After
applying Valiant's algorithm with \(E\), the following constraint theory is
learned:

\begin{align*}
    T=(\set{A,B,C}\lor\set{1,2,3})&\land(\set{A,B,C}\lor\set{1,2,4})\land\\
    (\set{A,B,C}\lor\set{1,3,4})&\land(\set{A,B,D}\lor\set{1,2,4})\land\\
    (\set{A,B,D}\lor\set{1,3,4})&\land(\set{A,B,D}\lor\set{2,3,4})\land\\
    (\set{A,C,D}\lor\set{1,2,4})&\land(\set{A,C,D}\lor\set{1,3,4})\land\\
    (\set{A,C,D}\lor\set{2,3,4})&\land(\set{B,C,D}\lor\set{1,2,4})\land\\
    (\set{B,C,D}\lor\set{1,3,4})&\land(\set{B,C,D}\lor\set{2,3,4})
\end{align*}

Our observation that the reject boxes for \(k=3\) is one-by-one is further
evidence by the fact that \textit{the learned constrained theory} \(T\)
\textit{is missing exactly four clauses, one clause for each observation in}
\(E\).

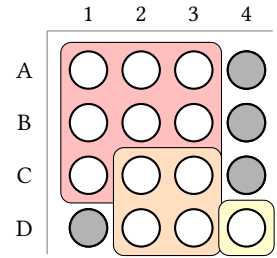
\begin{figure}[!h]
    \begin{tikzpicture}
        \coordinate (Origin)   at (0,0);
        \coordinate (XAxisMin) at (0.15,3.32);
        \coordinate (XAxisMax) at (3.15,3.32);
        \coordinate (YAxisMin) at (0.15,0.35);
        \coordinate (YAxisMax) at (0.15,3.32);
        \draw [thin, gray] (XAxisMin) -- (XAxisMax);
        \draw [thin, gray] (YAxisMin) -- (YAxisMax);
        \foreach \p in {1,2,3,4}{%
            \node[fill=none] at (0.7*\p,3.55) {\p};}
        \foreach \p\v in {1/D,2/C,3/B,4/A}{%
            \node[fill=none] at (-0.15,0.7*\p) {\v};}

        \foreach \x in {1,2,3,4}{%
            \foreach \y in {1,2,3,4}{%
                \node[thick,draw,circle,inner sep=5pt,fill=white!60,
                name=circle-\x-\y] at (0.7*\x,0.7*\y) {};}}

        \foreach \y in {2,3,4}{%
            \node[thick,draw,circle,inner sep=5pt,fill=gray!60,
            name=circle-\y] at (0.7*4,0.7*\y) {};}
        \node[thick,draw,circle,inner sep=5pt,fill=gray!60] at (0.7*1,0.7*1){};

        \begin{pgfonlayer}{bg}
            \node[draw, fill=red!25, rectangle, rounded corners, fit={(circle-1-2)
            (circle-3-4)}] {};
        \end{pgfonlayer}
        \begin{pgfonlayer}{bg}
            \node[draw, fill=orange!25, rectangle, rounded corners, fit={(circle-2-1)
            (circle-3-2)}] {};
        \end{pgfonlayer}
        \begin{pgfonlayer}{bg}
            \node[draw, fill=yellow!25, rectangle, rounded corners, fit={(circle-4-1)
            (circle-4-1)}] {};
        \end{pgfonlayer}
    \end{tikzpicture}
    \caption{Reject space for \(k=3\)---In addition to the observations
    rejected for \(k=1\) (shown in red) and \(k=2\) (shown in orange),
    observations \((D,4)\) is also rejected when \(k=3\) (shown in
    yellow).}\label{fig:proof_k3}
\end{figure}

In this setting, only one clause produces a reject box for a new observation
beyond the reject boxes produced by clauses at \(k=2\):
\((\set{A,B,C}\lor\set{1,2,3})\), with inverse \((\{D\}\land\{4\})\), as shown
in yellow in Figure~\ref{fig:proof_k3}. In this setting, \(k=n-1\) results in a
learned theory that ``overfits'' to the training set \(E\): \(T\) certifies
only observations from \(E\) an rejects all other observations. Thus, \(T\) is
\textit{rigid}, \ie{} it is \textit{least general}. Finally, we observe that
\(T\) certifies four samples and rejects twelve, \ie{} \(|\psi_3|=12\).

Given a domain $X$, which contains $X_i$, the set containing all possible
values for feature $i$, and a set of observations $\psi_k$ that are rejected by
a constraint theory $T_k$, which is a conjunction of clauses $t$, where $t$ is
a disjunction of literals $l_{k,i}$, which contain $k$ values of feature $i$,
we provide a general proof that $\psi_k\subseteq\psi_{k+1}$ (\ie{} $\forall e,
e\in\psi_{k}\implies e\in\psi_{k+1}$).

\begin{proof}
    Consider any $e\in\psi_k$, $1\leq k<n-1$, $n\geq3$. Without loss of
    generality we assume that every feature space $X_i$ is of size $n$. There
    must be some clause $t\in T_k$ that $e$ does not satisfy of the form:

    \[t=\bigvee_{i} l_{k,i} \]

    With

    \[ l_{k,i} \subset X_i \land | l_{k,i} | = k \land e_i \notin l_{k,i} \]

    Where $t$ describes the disjunction of literals $l_{k,i}$ of size $k$ that,
    for every feature $i$, contain a set of values that are within $X_i$, but
    do not contain the value $e_i$. Now, we can build $t'$, which is the clause
    $t$ with an additional value, $v_{k,i}$, in each literal:

    \[t'=\bigvee_{i} l_{k,i} \cup \{v_{k,i}\} \]

    With

    \[v_{k,i} \notin l_{k,i} \land v_{k,i} \subset X_i \land  e_i \neq v_{k,i}\]

    Where $t'$ is now clause $t$ with an additional value $v_{k,i}$ in each
    literal, where $v_{k,i}$ is within the set of all values for the feature
    $i$, but is not equal to $e_i$.

    Now, by construction, we have $t'\in T_{k+1}$, because literals are of size
    $k+1$ and the clause is strictly less constrained than $t$, since the
    literals of $t'$ can be satisfied by more samples than the literals of $t$.
    Since $t'\in T_{k+1}$, and we know that $t'$ rejects $e$ because $\forall
    i, e_i \notin l_{k,i}$ (\ie{} $t$ rejects $e$) and $\forall i, e_i \neq
    v_{k,i}$, which gives us $\forall i, e_i \notin l_{k+1,i}$ where $l_{k+1,i}
    = l_{k,i} \cup \{v_{k,i}$\}. Therefore, we have that $e \in \psi_{k+1}$. By
    induction, it can then be shown that:

    \[\psi_1\subseteq\cdots\subseteq\psi_k\subseteq\psi_{k+1}\subseteq\cdots\subseteq\psi_{n-1}\]
\end{proof}

\clearpage\section{Rationalizing Domain Constraints}\label{appendix-b}

\begin{figure}
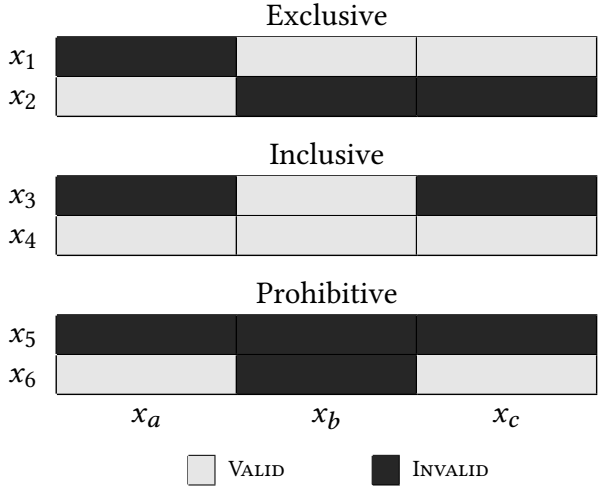

    \resizebox{0.95\columnwidth}{!}{%
    \begin{tabular}{r *{5}{|p{14.5mm}}}
        \multicolumn{1}{c}{} & \multicolumn{1}{c}{} &
        \multicolumn{1}{c}{Exclusive} & \multicolumn{1}{c}{} \\
        \hhline{~---}
        \(x_{1}\) & \cellcolor{oblack} & \cellcolor{owhite} &
        \cellcolor{owhite} \\
        \hhline{~---}
        \(x_{2}\) & \cellcolor{owhite} & \cellcolor{oblack} &
        \cellcolor{oblack} \\
        \hhline{~---}
        \noalign{\vspace{2mm}}
        \multicolumn{1}{c}{} & \multicolumn{1}{c}{} &
        \multicolumn{1}{c}{Inclusive} & \multicolumn{1}{c}{} \\
        \hhline{~---}
        \(x_{3}\) & \cellcolor{oblack} & \cellcolor{owhite} &
        \cellcolor{oblack} \\
        \hhline{~---}
        \(x_{4}\) & \cellcolor{owhite} & \cellcolor{owhite} &
        \cellcolor{owhite} \\
        \hhline{~---}
        \noalign{\vspace{2mm}}
        \multicolumn{1}{c}{} & \multicolumn{1}{c}{} &
        \multicolumn{1}{c}{Prohibitive} & \multicolumn{1}{c}{} \\
        \hhline{~---}
        \(x_{5}\) & \cellcolor{oblack} & \cellcolor{oblack} &
        \cellcolor{oblack} \\
        \hhline{~---}
        \(x_{6}\) & \cellcolor{owhite} & \cellcolor{oblack} &
        \cellcolor{owhite} \\
        \hhline{~---}
        \multicolumn{1}{c}{} & \multicolumn{1}{c}{\(x_{a}\)} &
        \multicolumn{1}{c}{\(x_b\)} & \multicolumn{1}{c}{\(x_c\)} \\
    \end{tabular}}

    \vspace{2.5mm}
    \resizebox{0.65\columnwidth}{!}{%
    \begin{tabular}{l|l|l|l|l}
        \hhline{~-~-~}
        {\hskip 10mm} & \cellcolor{owhite} & \textsc{Valid}{\hskip 10mm} &
        \cellcolor{oblack} & \textsc{Invalid} \\
        \hhline{~-~-~}
    \end{tabular}}

    \caption{Constraint Types --- An example visualization of three kinds of
    learned constraints between network protocols (\eg{} \(x_{1},
    x_{2},\dots,x_{6}\)) and services (\eg{} \(x_{a}, x_{b}\), \&
    \(x_{c}\))}\label{fig:ct}
\end{figure}

Here, we provide supplementary material on a conceptual model characterizing
the learned constraints by Valiant's algorithm into one of three types, as
shown in Figure~\ref{fig:ct}. Valiant's algorithm learns these three types of
constraints simultaneously; we found this conceptual model helping in
rationalizing the learned constraints.

\shortsection{Exclusive Constraints} Exclusive constraints are perhaps the most
intuitive types of relationships; they describe one-to-one mappings between two
variables. A constraint \(\mathscr{C}\) between \(X_{\alpha}\) and
\(X_{\beta}\) is said to be \textit{exclusive}, if:

\noindent\begin{equation*}
    \exists x_a \in X_{\beta}, \forall (x_1, x_2) \in X_{\alpha}^2, \mathscr{C}(x_1, x_a) \land
    \mathscr{C}(x_2, x_a) \implies x_1 = x_2
    \label{eq:xc}
\end{equation*}

\noindent where \(\mathscr{C}(x, y)\) is an indicator function that a
constraint exists between two variables \(x\) and \(y\). Thematically similar
to definitions for injective functions, exclusive constraints encode that if a
constraint exists between variables \(x_a\) and \(x_1\), then \(x_a\) does not
have a constraint for any other variable in the domain for which \(x_1\)
belongs to.  Conceptually, we can imagine such constraints between network
services and protocols. As an example, we can expect that some services, such
as \texttt{SSH}, can only be used with \texttt{TCP}.

\shortsection{Inclusive Constraints} Unlike exclusive, inclusive constraints
describe one-to-many mappings. Again, a constraint \(\mathscr{C}\) between
\(X_{\alpha}\) and \(X_{\beta}\) is said to be \textit{inclusive}, if:

\noindent\begin{equation*}
    \exists x_a \in X_{\beta}, \forall (x_1, x_2) \in X_{\alpha}^2, \mathscr{C}(x_1, x_a) \land
    \mathscr{C}(x_2, x_a) \centernot\implies x_1 = x_2
    \label{eq:ic}
\end{equation*}

\noindent In other words, it is \textit{not} necessary for \(x_1\) and \(x_2\)
to be the same variable if there exists constraints \(\mathscr{C}(x_1, x_a)\)
and \(\mathscr{C}(x_2, x_a)\). We can again imagine a scenario where a service,
for example \texttt{NTP}, can be used with multiple protocols, such as
\texttt{TCP} and \texttt{UDP}.

\shortsection{Prohibitive Constraints} We call the final constraint type
prohibitive constraints. These constraints describe regions for which no
observation can exist\footnotemark. We formalize a prohibitive constraint
\(\mathscr{C}\) between \(X_{\alpha}\) and \(X_{\beta}\) as:

\footnotetext{There are constraint learning approaches that leverage
\textit{negative examples}, which are realizations of inputs that cannot exist.
Practical datasets used for machine learning do not provide such examples, and
thus prohibitive constraints give us the flexibility to \textit{infer} negative
examples from gaps between values in positive examples.}

\noindent\begin{equation*}
    \forall x_1 \in X_{\alpha}, \nexists x_a \in X_{\beta} ~such ~that ~\mathscr{C}(x_1, x_a) = 1
    \label{eq:pc}
\end{equation*}

\noindent Prohibitive constraints are especially interesting. At first, they
seem redundant (one could consider invalid regions to simply be the dual of
valid regions) or non-informative (a variable in a dataset that has no
observation surely cannot be useful to any learning algorithm). The necessity
of prohibitive constraints is rooted in learning relationships with variables
that live in a \textit{continuous} domain. While such variables can take any
real value in theory, there exists real-world phenomena for which continuous
variables instead take values that can be approximated as discrete clusters of
values. Prohibitive constraints grant us the flexibility to learn such
phenomena.

As an example from networks, consider packet sizes; it is a well-known
phenomena that packet sizes on the Internet closely follow a bimodal
distribution~\cite{thompson_wide-area_1997, williamson_internet_2001} (\ie{}
packet lengths are either small or large). Consider this observation at the
extreme, that is, let the two modes be non-overlapping. In a practical setting,
this means that the \textit{largest} packet from the ``small'' distribution is
smaller than the \textit{smallest} packet from the ``large'' distribution.
Variables that describe continuous phenomena can exhibit these regions and
prohibitive constraints are necessary to model these contexts.

\clearpage\section{Miscellany}\label{appendix-c}

\begin{table}[ht]
    \begin{tabular}{ll}
        \toprule
        Dataset & Features \\
        \midrule
        \underline{\nsl{}} & 1.\texttt{Flag} \\
        & 2.\texttt{Src Bytes}\\
        & 3.\texttt{Dst Bytes}\\
        & 4.\texttt{Land}\\
        & 5.\texttt{Num Compromised}\\
        & 6.\texttt{Srv Serror Rate}\\
        & 7.\texttt{Rerror Rate}\\
        & 8.\texttt{Diff Srv Rate}\\
        & 9.\texttt{Srv Diff Host Rate}\\
        & 10.\texttt{Dst Host Srv Serror Rate}\\
        & 11.\texttt{Dst Host Rerror Rate}\\
        \midrule
        \underline{\ph{}} & 1.\texttt{UrlLength}\\
        & 2.\texttt{NumNumericChars}\\
        & 3.\texttt{NumSensitiveWords}\\
        & 4.\texttt{PctExtHyperlinks}\\
        & 5.\texttt{PctNullSelfRedirectHyperlinks}\\
        & 6.\texttt{FrequentDomainNameMismatch}\\
        & 7.\texttt{SubmitInfoToEmail}\\
        & 8.\texttt{PctExtResourceUrlsRT}\\
        & 9.\texttt{ExtMetaScriptLinkRT}\\
        & 10.\texttt{PctExtNullSelfRedirectHyperlinksRT}\\
        \bottomrule
    \end{tabular}
    \caption{Features}\label{tab:features}
\end{table}

\subsection{Hyperparameters}\label{appendix-c:hyperparameters}

\begin{table}[ht]
    \resizebox{\columnwidth}{!}{%
    \begin{tabular}{lccc}
        \toprule
        Dataset & \# Neurons per Hidden Layer & Learning Rate & Iterations \\
        \midrule
        \nsl{} & 60, 32  & \(10^{-3}\) & \SI{16}{} \\
        \ph{}  & 20 & \(10^{-2}\) & \SI{15}{} \\
        \bottomrule
    \end{tabular}}
    \caption{Hyperparameters}\label{tab:hyperparameters}
\end{table}

For our models, we use multilayer-perceptrons using \textsc{ADAM} optimizer.
Hyperparameters are shown in Table~\ref{tab:hyperparameters}.

\subsection{Table of Symbols}\label{appendix-c:symbols}

\begin{table}[h]
    \begin{tabular}{ll}
        \toprule
        \textbf{\textit{Symbol}}      & \textbf{\textit{Meaning}}\\
        \midrule
        \multicolumn{2}{l}{\underline{\textbf{\textit{Universal}}}}\\
        \(e\)                & sample or observation\\
        \(E\)                & dataset or collection of observations\\
        &\\
        \multicolumn{2}{l}{\underline{\textbf{\textit{Machine Learning}}}}\\
        \(e^*\)              & adversarial example\\
        \(y\)                & sample label\\
        \(\theta\)           & model parameters\\
        \(f_\theta\)         & model with parameters \(\theta\)\\
        \(\jacobian\)        & Jacobian of a model\\
        \textbf{S}           & Saliency Map\\
        \(\hat{y}\)          & model prediction\\
        \(\alpha\)           & perturbation magnitude\\
        \(\mathscr{p}\)      & parameter for some \lp{\mathscr{p}}-norm\\
        \(\phi\)             & budget (measured as a distance)\\
        \(\mathcal{B}_\phi\) & norm-ball of radius \(\phi\)\\
        &\\
        \multicolumn{2}{l}{\underline{\textbf{\textit{Formal Logic}}}}\\
        \(\vdash\)           & logical entailment\\
        \(\in\)              & set membership\\
        \(X\)                & domain\\
        \(X_i\)              & \(i\)th feature (or variable) space\\
        \(x_i\)              & some value for feature \(i\)\\
        \(C\)                & possible constraints from a domain\\
        \(t\)                & a clause in a constraint theory\\
        \(T\)                & a constraint theory\\
        \(H\)                & a constraint theory with partial assignments\\
        \(\psi\)             & set of observations rejected by a constraint theory\\
        \(P\)                & Pseudo-power set\\
        \(p\)                & set within some pseudo-power set\\
        \(\Xi\)              & space of possible observations\\
        \(\Lambda\)          & constraint-compliant set of observations\\
        \(\Psi\)             & constraint-noncompliant set of observations\\
        \bottomrule
    \end{tabular}
    \caption{Symbols used in this paper}\label{tab:symbols}
\end{table}
\subsection{Dataset Details}\label{appendix-c:dataset_details}

Table~\ref{tab:features} shows the features used in our datasets after feature
reduction. Details on the meaning behind the features can be found
in~\cite{tavallaee_detailed_2009} for the \nsl{} and in~\cite{chiew_new_2019}
for \ph{}.

\subsection{DPLL}\label{appendix-c:dpll}

Constraint learning is a historical problem in computer science and shares many
parallels with satisfiability problems. While constraint learning tries to
learn what the constraints are, satisfiability attempts, as the name suggests,
to return an assignment that satisfies a set of boolean expressions.

For constraint satisfaction, we use \textsc{Davis-Putnam-Logeman-Loveland}
(\dpll{})~\cite{davis_computing_1960}, shown in Algorithm~\ref{alg:dpll}.
\dpll{} has some characteristics that make it ideal for our task, namely: (1)
it accepts boolean formulae in CNF, which is the native form of the constraint
theories learned by Valiant's Algorithm, and (2) it is a
\textit{backtracking-based} search algorithm. \dpll{} iteratively builds
candidate solutions for a given expression, which is a property we exploit,
detailed in Section~\ref{approach:projecting_adversarial_examples}.

\begin{algo}[t]
    \SetAlgoLined{}
    \KwIn{set of boolean clauses \(H\)}
    \KwOut{a truth value}
    \If{\(H\) contains a contradiction}{%
        \Return{\textsc{False}}
    }
    \If{\(H\) contains an assignment for every variable}{%
        \Return{\textsc{True}}
    }
    \For{\(C^* \in \set{C \mid C \in H \land |C| = 1}\)}{%
        \(H \leftarrow \texttt{UnitPropagate}(C, H)\)
    }
    \For{\(l^* \in \set{l \mid l \in H \land \texttt{Pure}(l)}\)}{%
        \(H \leftarrow \texttt{PureLiteralElimination}(l^*, H)\)
    }
    \(l \leftarrow\) arbitrarily select an unassigned literal \\
    \If{\(\dpll{}(H \cup \set{l \leftarrow \textrm{\textsc{True}}}) =
    \textrm{\textsc{True}}\)}{%
        \Return{\textsc{True}}
    }
    \Else{%
        \Return{\(\dpll{}(H\ \cup \set{l \leftarrow
        \textrm{\textsc{False}}})\)}
    }
    \caption{Davis-Putnam-Logemann-Loveland}%
    \label{alg:dpll}
\end{algo}

\dpll{} frames constraint satisfaction as a search problem; first initialized
with a set of boolean clauses \(H\) containing unassigned literals, the
algorithm first assigns a literal \(l\) to either \textsc{True} or
\textsc{False}, and recursively calls itself with the new assignment for \(l\).
\dpll{} returns \textsc{False} if a contradiction is reached or \textsc{True}
if all literals are assigned. \dpll{} has a runtime advantage over other
backtracking algorithms as it performs two simplifications to \(H\) at each
call: \texttt{UnitPropagate} and \texttt{PureLiteralElimination}.
\texttt{UnitPropagate} assigns values to literals who are the only members of
their clauses (as only one assignment makes such clauses true).
\texttt{PureLiteralElimination} assigns the necessary value to literals who are
\textit{pure}, that is, a literal \(l\) is either \(l\) or \(\lnot l\) for all
clauses in \(H\). Therefore, such literals can be assigned so that all clauses
containing them are true.

\subsection{Constraint Representation}\label{appendix-c:constraint_representation}

Recall Valiant's algorithm; it determines if an observation \(e\) complies with
a constraint theory \(T\) by iterating over all clauses \(t\in T\) and
evaluating if at least one literal in \(t\) is satisfied by the feature values
of \(e\).  Depending on the cardinality of each literal, determining if a
clause is satisfied through a linear search could be intractable. Instead, we
use a set-based representation for literals as this reduces clause satisfaction
to be on the order of the number of literals in the clause (since
set-membership queries can be done in constant time).

While this optimization is appropriate for boolean and categorical features,
continuous features require a different representation to be efficient. Recall
that we leverage \optics{} to cluster feature values into discrete bins
representing sets of ranges (\eg{} \(\set{x\mid (0.25 \leq x < 0.50) \lor (0.75
\leq x < 1.00)}\)). Thus, checking if clauses are satisfied with continuous
features can be done in \(\bigo{}(\log (n))\) time with binary search, where
\(n\) describes the number of bins.

In practice these optimizations yielded a tractable constraint learning
process. The \nsl{} constraint generation process executed in just over 3 days,
and the phishing dataset completed in about 1 day (with \optics{} consuming the
vast majority of time, detailed in Section~\ref{evaluation:scalability}). Note
that learning need only be executed once for a set of training data.  Moreover,
there are algorithmic optimizations that could enable the integration of new
data in an incremental way~\cite{muggleton_optimal_1993}. We will explore these
techniques in future work.

\end{document}